\documentclass[12pt]{article}         
\usepackage{amsfonts}      
    
\newcommand{\be}{\begin{equation}}
\newcommand{\ee}{\end{equation}}
\newcommand{\ba}{\begin{array}}
\newcommand{\ea}{\end{array}}
\newcommand{\bea}{\begin{eqnarray}}
\newcommand{\eea}{\end{eqnarray}}
\newcommand{\beaa}{\begin{eqnarray*}}
\newcommand{\eeaa}{\end{eqnarray*}}
\newcommand{\al}{\alpha}

\newcommand{\de}{\delta}

\newcommand{\ep}{\epsilon}
\newcommand{\ga}{\gamma}
\newcommand{\Ga}{\Gamma}
\newcommand{\la}{\lambda}

\newcommand{\si}{\sigma}
\renewcommand{\th}{\theta}

\newcommand{\rb}{\right]}
\newcommand{\lb}{\left[}

\newcommand{\bl}{\biggl(}
\newcommand{\br}{\biggr)}

\renewcommand{\(}{\left(}
\renewcommand{\)}{\right)}
\newcommand{\ato}{a_2^{(1)}}
\newcommand{\att}{a_2^{(2)}}
\newcommand{\uq}[1]{$U_q(#1)$}
\newcommand{\nn}{\nonumber}
\newcommand{\fns}{\footnotesize}
\newcommand{\scs}{\scriptsize}
\newcommand{\shs}{\shortstack}
\newcommand{\ol}{\overline}
\newcommand{\ul}{\underline}
\newcommand{\noi}{\noindent}
\newcommand{\hs}{\hspace}
\newcommand{\vs}{\vspace}
\newcommand{\lra}{\longrightarrow}
\newcommand{\ot}{\otimes}

\newcommand{\cA}{{\cal A}}
\newcommand{\cB}{{\cal B}}

\newcommand{\cD}{{\cal D}}

\newcommand{\mup}{\mu^{\prime}}   
  
\newcommand{\tK}{\tilde{K}}

\advance\textheight by \topskip
\makeatletter

\@addtoreset{equation}{section}
\def\section{\@startsection {section}{1}{\z@}{-8.5ex plus -1ex minus
 -.2ex}{3.3ex plus .2ex}{\large\bf\centering}}
\def\subsection{\@startsection{subsection}{2}{\z@}{-3.25ex plus
 -1ex minus -.2ex}{1.5ex plus .2ex}{\bf}}
\def\subsubsection{\@startsection{subsubsection}{3}{\z@}{-3.25ex plus%
 -1ex minus -.2ex}{1.5ex plus .2ex}{\sl}}

\begin{document}        

\newpage  
\begin{titlepage}  
\begin{flushright} 
DTP--98/29 \\ 
hep-th/9806003 \\  
{\bf (revised version: July 1998)}  
\end{flushright}  
\vspace{1cm}  
\begin{center}  
{\Large {\bf On $\ato$ Reflection Matrices and\\  \vs{0.5cm}
 Affine Toda Theories}}\\
\vspace{1.5cm}  
{\large Georg M.\ Gandenberger}\footnote{\noi E-mail:  
G.M.Gandenberger@durham.ac.uk}\\   
\vs{0.5cm}  
{\em Department of Mathematical Sciences\\  
Durham University\\   
Durham DH1 3LE, U.K.}\\  
\vspace{2cm}   
{\bf{ABSTRACT}}  
\end{center}  
\begin{quote}  
We construct new non-diagonal solutions to the boundary
Yang--Baxter--Equation corresponding to a two-dimensional field
theory with \uq{\ato} quantum affine symmetry on a half-line.    
The requirements of boundary unitarity and boundary crossing symmetry
are then used to find overall scalar factors which lead to
consistent reflection matrices. Using the boundary bootstrap equations
we also compute the reflection factors for scalar bound states
(breathers). These breathers are expected to be identified with the 
fundamental quantum particles in  $\ato$ affine Toda field theory and
we therefore obtain a conjecture for the affine Toda reflection factors.
We compare these factors with known classical results and discuss
their duality properties and their connections with particular boundary
conditions.  
\end{quote}  
  
\vfill  
  
\end{titlepage}

\section{Introduction} 

There has been an increasing amount of effort towards the
understanding of two-dimensional integrable field theories with
reflecting boundaries in recent years. Apart from their role as toy
models for higher dimensional field theories and their connections
with open string theories, these models  have also found some 
applications in real physical situations like impurity problems in
condensed matter physics.  
Many interesting open questions remain in the classical as well as
in the quantum theory of integrable models with reflecting
boundaries. 

Classically, one of the first questions to ask is whether there are
any boundary conditions for a given integrable field theory which 
preserve integrability. This problem was extensively studied in a
series of papers \cite{corri94, bowco95} (for a recent review
see \cite{corri96}) in which the authors 
examined real coupling affine Toda field theories (ATFT) on a
half-line ($-\infty < x \leq 0$), defined by a Lagrangian of the form
\be
{\cal L}_B = \Theta(-x){\cal L} - \delta(x){\cal B}(\phi)\;,
\ee
in which ${\cal L}$ is the usual affine Toda Lagrangian and ${\cal B}$
is some boundary potential, which is chosen to depend only on the
field $\phi$ and not on any time derivatives of $\phi$.
In particular, for the case of $a_n^{(1)}$ 
ATFTs it was found that ${\cal L}_B$ defines an integrable theory if
the boundary potential takes the following form 
\be
{\cal B} = \sum_{j=0}^n b_je^{\al_j \frac{\phi}2}\;, \label{todaB}
\ee
in which the $\al_j$ are the simple roots of $a_n^{(1)}$ and $b_j$
are some boundary parameters. Surprisingly, it turned out that only 
in the simple case of the sinh--Gordon theory, which is $a_1^{(1)}$
ATFT, the boundary parameters are free and hence there are an infinite
number of integrable boundary conditions. Whereas 
in all other cases ($n \geq 2$) only a finite number of
integrable boundary potentials exist. 
For instance for the $\ato$ ATFT on a half-line it was found that
there are only nine possible boundary terms which lead to a
classically integrable theory, namely  
\bea  
\mbox{either} && \;\;\;\;\; b_0 = b_1 = b_2 = 0\;, \nn \\
\mbox{or} && \;\;\;\;\; b_j = \pm 2 \;\;\;\;\;\;\;\;\;\;\;\;\;\;
(j=0,1,2)\;. \nn 
\eea 

In the quantum case all on-shell information of a
two-dimensional integrable field theory is
contained in its $S$-matrix. In the presence of a reflecting boundary
the $S$-matrix has to be supplemented by a so-called reflection matrix,
which describes the scattering of a particle from the boundary. 
Following earlier works by Cherednik \cite{chere84} and Sklyanin
\cite{sklya88} on factorisable scattering on a half-line, 
in a seminal paper \cite{ghosh94} Ghoshal and Zamolodchikov  studied
the general requirements imposed by integrability on the reflection 
matrices. 
In particular they were able to solve the
boundary Yang--Baxter--Equation and construct reflection matrices for
the sine--Gordon theory on a half-line. 
One of the most difficult problems in the boundary sine-Gordon theory
turned out to be the question of how to relate the free parameters
appearing 
in the reflection matrices with the boundary parameters
in the Lagrangian. This is still an open problem, even though
some progress has been achieved in a recent paper
\cite{corri97}, in which it was noticed that the breather reflection
factors of sine--Gordon theory \cite{ghosh94b} should be identical to
the reflection factors for the particles in sinh--Gordon. Hence these
reflection factors could be compared with results from perturbation
theory.  

On the other hand, since in any of the higher order $a_n^{(1)}$ ATFTs
only a finite number of integrable boundary conditions exist,  
it would seem feasible that, if we could find the general form of the  
reflection factors for these theories, it should be fairly 
straightforward to associate them with particular boundary
conditions. However, despite several attempts the construction of
reflection factors for ATFTs on a half-line has been elusive so far.    

Another interesting problem is the question of duality. The
real affine Toda $S$-matrices were found to display an intriguing
weak--strong coupling duality \cite{brade90}. Up to now it has been
unclear whether a similar duality will be present in the reflection
matrices of ATFTs.  However, recently it was suggested in
\cite{corri97} that in the sinh--Gordon theory duality relates the
reflection factors of the Neumann boundary condition ($b_0=b_1=0$)
with those of the `positive' boundary condition ($b_0=b_1=+2$). It
would be interesting to see wether a similar relation could be true
for any other ATFT.

One of the main results in this paper is a derivation of the
reflection factors for the simplest ATFT with more than one particle,
namely the theory based on the algebra $\ato$. This derivation uses an
analogy with 
the $S$-matrices for ATFTs on the whole line. It has been realised  
over the past few years that the real Toda $S$-matrices can be
constructed using exact $S$-matrices with quantum affine symmetries
\cite{gande95, gande95b, gande96, gande98}. These \uq{\hat{g}}
invariant $S$-matrices, which were 
constructed using trigonometric $R$-matrices, have been 
conjectured to describe the 
scattering of the solitons in imaginary coupled ATFTs. It then turned
out that the scalar bound states (breathers) in these theories can be
identified with the fundamental Toda quantum particles and that the
scattering amplitudes for the breathers are identical to the real Toda
$S$-matrices after analytic continuation of the coupling constant.  
Here we introduce a similar construction for a \uq{\ato} invariant
theory with reflecting boundary.
 
The layout of this paper is as follows. Section 2 briefly reviews the 
\uq{\ato} invariant \mbox{$S$-matrix} and the use of the 
Faddeev--Zamolodchikov algebra. Section 3 provides an introduction to 
the boundary Yang--Baxter--Equation and we construct solutions related
to the \uq{\ato} invariant \mbox{$S$-matrix}.  
In section 4 we discuss the restrictions imposed on these solutions by
boundary unitarity and boundary crossing, and we construct suitable
overall scalar factors.    
\mbox{Section 5} examines the implications of the boundary bootstrap 
conditions, and we use the bootstrap equations in order
to compute reflection factors for the breathers. 
Finally in section 6 we give a conjecture for the reflection factors
of real coupling $\ato$ ATFT on a half-line and discuss their duality
properties.   
Two appendices provide some of the computational details which have
been omitted from the main text.

\section{The \uq{\ato} $S$-matrix}    

This is a brief review of the \uq{\ato} invariant trigonometric  
$S$-matrix which was studied in detail in \cite{gande95, gande96}.   
We consider a two-dimensional integrable field theory which displays 
a \uq{\ato} affine quantum symmetry. The theory contains two 
multiplets transforming under the two fundamental representations of 
\uq{\ato}. Each of these multiplets contains three states, which 
we label by $A_1, A_2, A_3$ and $\ol A_1   
,\ol A_2,\ol A_3$, respectively.   
The $S$-matrix of the theory is an intertwining map on the two 
representation spaces (denoted by $V_1$ and $V_2$) 
\be 
S_{a,b}(\th): V_a \otimes V_b \lra V_b \otimes V_a\;, 
\;\;\;\;\;\;\;\;\; (a,b=1,2)\;, 
\ee 
in which $\th$ is the rapidity difference of the incoming states.  
Instead of the $\th$, from now on we will always use the 
following parametrisation of the rapidity variable:    
\be   
\mu = -i\frac{3\la}{2\pi}\th\;,   
\ee 
in which $\la$ is a coupling constant parameter.  
Due to the quantum affine symmetry of the theory, the $S$-matrix can be 
constructed using the \uq{\ato} $R$-matrix in the principal gradation. 
This construction was first performed in \cite{hollo93} (for more 
details see \cite{gande96}).  
 
Rather than using the $R$-matrix, for our purposes it will prove more 
convenient to write the scattering theory in terms of a 
non-commutative algebra, the so called Faddeev--Zamolodchikov (FZ) 
algebra. The generators of this algebra are denoted by $A_i(\mu)$, 
$\ol A_i(\mu)$, which represent the fundamental states in the theory, 
and the possible scattering processes are expressed in terms of the 
following braiding relations:   
\bea    
A_j(\mu_1)A_j(\mu_2) &=& S^I(\mu_{12}) A_j(\mu_2)A_j(\mu_1)\;, \nn    
\\     
A_j(\mu_1)A_k(\mu_2) &=& S^T(\mu_{12}) A_k(\mu_2)A_j(\mu_1) +    
S^{R(j,k)}(\mu_{12}) A_j(\mu_2)A_k(\mu_1)\;,\;\;\;\;\; (j\neq 
k) \label{FZAi}     
\eea    
in which $\mu_{12} = \mu_1 -\mu_2$, and the elements of the $S$-matrix 
are given as      
\bea     
S^I(\mu) & = & F(\mu)\;,     
\nn \\     
S^T(\mu) & = & \frac{\sin(\pi\mu)}{\sin(\pi(\la-\mu))} F(\mu)\;,    
\nn \\     
S^{R(j,k)}(\mu) & = & \frac{\sin(\pi\la)}{\sin(\pi(\la -\mu))}    
e^{\nu(j,k)i\pi\frac{\mu}3}F(\mu)\;,     
\eea    
\vs*{8pt}     
\hs{2.5cm} in which 
\be
\nu(j,k) \equiv \left\{    
\begin{array}{ll}+1, \hs{5pt}    
& \mbox{if  $(j,k) = (1,2),(2,3)$ or $(3,1)$}\;, \\ -1, &    
\mbox{if $(j,k) = (2,1),(3,2)$ or $(1,3)$}\;.    
\end{array}    
\right. \label{nukl}
\ee    
The overall scalar factor $F(\mu)$ can be written as an 
infinite product of Gamma functions   
\be  
F(\mu) = - \prod_{j=1}^{\infty}  
\frac{\Ga(\mu+3j\la-2\la+1)   
\Ga(\mu+3j\la-\la)\Ga(-\mu+3j\la-3\la+1) \Ga(-\mu+3j\la)}  
{\Ga(-\mu+3j\la-2\la+1)\Ga(-\mu+3j\la-\la)\Ga(\mu+3j\la-3\la+1) 
\Ga(\mu+3j\la)}\;.  \label{F11} 
\ee  
Scattering processes involving the states $\ol A_i$ can be easily 
derived from (\ref{FZAi}) using crossing symmetry, e.g. 
\bea    
\ol A_j(\mu_1) A_j(\mu_2) &=& S^I(\frac 32 \la - \mu_{12}) 
A_j(\mu_2)\ol A_j(\mu_1) + S^{R(j,k)}(\frac 32 \la - \mu_{12}) 
A_k(\mu_2)\ol A_k(\mu_1)\;, \nn   \\     
\ol A_k(\mu_1) A_j(\mu_2) &=& S^T(\frac 32 \la - \mu_{12}) A_j(\mu_2) 
\ol A_k(\mu_1)\;.  
\eea    
 
This $S$-matrix has been conjectured to describe 
the scattering of solitons in $\ato$ ATFT with imaginary coupling 
constant. In the following we will therefore often call the 
states $A_i$, $\ol A_i$ solitons. Apart from these fundamental 
solitons there are also bound states in the theory, which correspond 
to simple poles in the $S$-matrix. There are two kinds of bound states 
in the theory, namely scalar bound states or breathers and excited
solitons.   
Here we are only interested in the lowest mass breathers, i.e.\ those 
bound states corresponding to the pole $\mu = \frac 32 \la - 1$ which 
appears in the cross channel of $S^I(\mu)$.  
In order to describe the scattering of these bound states we  
have to extend the FZ algebra by generators $B_1(\mu)$ and 
$\ol B_1(\mu)$ corresponding to the lowest breathers. It has been 
shown in \cite{gande95} that these generators can be defined formally 
as   
\bea 
B_1(\frac{\mu_1+\mu_2}2) &=& \lim_{\mu_2-\mu_1 \to 
\frac 32 \la - 1}\,\sum_{m=1}^3\al_m\, A_m(\mu_1)\, \ol 
A_m(\mu_2)\;, \nn \\ 
\ol B_1(\frac{\mu_1+\mu_2}2) &=& \lim_{\mu_2-\mu_1 \to 
\frac 32 \la - 1}\,\sum_{m=1}^3\al_m\, \ol A_m(\mu_1)\, 
A_m(\mu_2)\;, \label{FZB}
\eea 
in which  
\[ 
\al_1 = e^{i\pi\frac 23}\;, \;\;\;\; \al_2 = 1\;, 
\;\;\;\; \al_3 = e^{-i\pi\frac 23}\;. 
\]  
The scattering of two of these breathers with each other can then 
be described by the braiding relation 
\[
B_1(\mu_1)B_1(\mu_2) = S_{B_1,B_1}(\mu_{12}) B_1(\mu_2)B_1(\mu_1)\;, 
\] 
and similarly for the conjugate breathers $\ol B_1$.  
One can compute the breather scattering amplitudes by using the
bootstrap equations, and it was found that 
\be 
S_{B_1,B_1}(\mu) = S_{\ol B_1,\ol B_1}(\mu) = \bl 1 \br\, \bl \la 
\br\, \bl -1-\la \br\;, \label{SBB} 
\ee 
and  
\[ 
S_{B_1, \ol B_1}(\mu) = S_{\ol B_1,B_1}(\mu) = S_{B_1,B_1}(\frac 32 
\la- \mu)\;, 
\] 
in which we have used the bracket notation 
\be 
\bl a \br \equiv \frac{\sin(\frac{\pi}{3\la}(\mu+a))} 
{\sin(\frac{\pi}{3\la}(\mu-a))}\;. \label{bracket1} 
\ee  
 
Finally we note that if we choose the parameter $\la$ to be related to
the affine Toda coupling constant $\beta$ in the following
way\footnote{Note, that this is not really a choice, but can be
derived by considering non-local conserved charges in imaginary
coupled ATFT.}   
\be 
\la = \frac{4\pi}{\beta^2}-1\;,  \label{labeta}
\ee 
then it turns out that after analytic continuation, $\beta \to 
i\beta$, the breather scattering amplitudes (\ref{SBB}) are 
identical to the $S$-matrix for real coupling $\ato$ ATFT 
\cite{brade90}. This  
establishes the fact that, just as in the sine--Gordon theory, the 
lowest mass breathers can be identified with the fundamental quantum 
particles in $\ato$ ATFT. This identification will later provide the 
basis for our conjecture of the reflection factors in   
$\ato$ ATFT on a half-line.

\section{The Boundary Yang--Baxter--Equation}     

From now on we will deal with a \uq{\ato} invariant scattering theory in 
 which the spatial coordinate is restricted to the half-line $-\infty 
 < x \leq 0$.   
We assume that this theory is still integrable and that any 
scattering processes far away from the boundary are still 
described by the \uq{\ato} invariant $S$-matrix from the previous 
section. Incoming states moving towards the boundary at $x=0$ will
 be reflected  into outgoing states with negative rapidity. This 
 reflection of states from the boundary will be described by 
 so-called reflection matrices.   
  
First of all, we will assume that the reflection of solitons is always  
multiplet changing, which means an incoming state of type $A_i$ can 
only reflect into a state of type $\ol A_j$ and vice versa\footnote{For 
the general case of a $a_n^{(1)}$ theory we expect that solitons in the 
$a$th multiplet reflect into solitons in the charge conjugate 
$(n+1-a)$th multiplet.}. Hence the reflection matrices are maps on the 
representation spaces of the form    
\bea    
K(\mu): V_1 &\lra & V_{2}\;, \nn \\  
\ol K(\mu): V_2 &\lra & V_{1} \;.   \label{Kmatrices} 
\eea    
This assumption of multiplet changes for the reflection of solitons 
may seem rather unmotivated at this stage. However, some recent work
by Gustav Delius \cite{deliu98} on classical soliton solutions in
ATFTs on a half-line 
has shown that the only classical solutions possible are those in
which a soliton is reflected into a corresponding antisoliton. 
Furthermore, this property will also lead to the fact that the 
breathers do not change after reflection, which is expected if we want 
to identify the lowest breathers with the fundamental quantum 
particles in ATFT.  
In appendix A we will briefly mention other known solutions to the
BYBE which are not of the type (\ref{Kmatrices}).
 
The main restriction on the explicit form of the reflection matrices 
(\ref{Kmatrices}) comes from the  
boundary Yang--Baxter--Equation (BYBE), which is the analogue of the 
Yang--Baxter--Equation (YBE) in scattering theories on the full line.  
In our case the BYBE takes the following form:    
\bea    
\lefteqn{[I_2 \ot K(\mup)]\,.\,S_{1,2}(\mu+\mup)\,.\,[I_1 \ot   
K(\mu)]\,.\, S_{1,1}(\mu-\mup) =} \hs{4cm} \nn \\      
& & = S_{2,2}(\mu-\mup) \,.\,[I_{2} \ot K(\mu)] \,.\,    
S_{1,2}(\mu+\mup) \,.\,[I_1 \ot K(\mup)]\;,    
\eea     
in which $I_a$ denotes the identity on $V_a$ and both sides map $V_1 
\ot V_1$ into $V_{2}\ot V_{2}$.   
The corresponding equation for $\ol K(\mu)$ is the same with all 
indices 1 and 2 exchanged.  
 
In order to explicitly solve these equations it is useful to write 
them in terms of the FZ algebra. Since $V_1$ and $V_2$ are three 
dimensional, we can write the reflection matrices as $3\times3$ 
matrices $(K_i^j)_{i,j=1,\dots,3}$. We then extend the FZ algebra 
(\ref{FZAi}) for the theory on a half-line by the following 
relations:  
\bea  
A_j(\mu)\, \cB &=&  K_j^k(\mu) \ol A_k(-\mu)\,\cB\;, \nn \\  
\ol A_j(\mu)\, \cB &=&  \ol K_j^k(\mu) A_k(-\mu)\,\cB\;, 
\label{FZrefl}  
\eea  
in which $\cB$ denotes the boundary at $x=0$ and summation over 
repeated indices is assumed.  
{\em Figure 1} shows the diagrammatic representation of these simple 
soliton reflection processes. (In these and all following diagrams
time is meant to be increasing up the page.) 
%
%
%
\begin{center} 
\begin{picture}(350,180)(-10,-30) 
\put(50,10){\rule{3pt}{120pt}}
\put(50,70){\line(-1,-1){45}} 
\put(50,70){\line(-1,1){45}} 
\put(65,67){\shs{$= K_j^k(\mu)\;,$}}
\put(-5,15){\shs{\fns{$A_j$}}} 
\put(-2,120){\shs{\fns{$\ol A_k$}}} 
\put(300,10){\rule{3pt}{120pt}}
\put(300,70){\line(-1,-1){45}} 
\put(300,70){\line(-1,1){45}} 
\put(315,67){\shs{$= \ol K_j^k(\mu)\;.$}}
\put(245,15){\shs{\fns{$\ol A_j$}}} 
\put(248,120){\shs{\fns{$A_k$}}} 
\put(100,-30){\shs{\em Figure 1: Reflection matrices}}
\end{picture} 
\end{center} 

We can also write the $S$-matrix in explicit matrix form, such
that the element $S_{i,j}^{k,l}(\mu_{12})$ denotes the scattering
amplitude for the scattering process $A_i(\mu_1) +A_j(\mu_2) \lra   
A_k(\mu_2)+A_l(\mu_1)$. The braiding relations (\ref{FZAi}) can then 
be written in a more compact form: 
\be 
A_j(\mu)A_k(\mu) = S_{j,k}^{l,m}(\mu_{12}) 
A_l(\mu_2)A_m(\mu_1)\;, 
\ee 
in which the non-zero components of the $S$-matrix are 
\bea    
S_{i,i}^{i,i}(\mu) &=& S^I(\mu)\;,     
\nn \\    
S_{i,j}^{j,i}(\mu) &=& S^T(\mu)\;,     
\nn \\    
S_{i,j}^{i,j}(\mu) &=& S^{R(i,j)}(\mu)\;, \;\;\;\;\;\;\;  
(\mbox{for}\;\;\;  i,j  = 1, \dots, 3;\;\;\;\; \mbox{and}\;\;\;\;\; i 
\neq j) \;.  \label{Smcomp}  
\eea    
The matrix elements involving the scattering of antisolitons can again 
be obtained easily using crossing symmetry: 
\[    
S_{i,\ol j}^{\ol k,l}(\mu) = S_{k,i}^{l,j}(\frac 32 \la -\mu)\;, 
\;\;\;\;\;\;\; \mbox{and} \;\;\;\;\;\;\;\;  S_{\ol i,\ol j}^{\ol k,\ol
l}(\mu) = S_{l,k}^{j,i}(\mu)\;. 
\] 
We can therefore write the BYBE as a matrix equation  
\be  
K_j^k(\mup)\, S_{i,\ol k}^{\ol l,m}(\mu+\mup)\, 
K_m^n(\mu)\, S_{\ol l,\ol n}^{\ol p,\ol r}(\mu-\mup) =  
S_{i,j}^{k,l}(\mu-\mup)\, K_l^m(\mu)\, S_{k,\ol m}^{\ol 
p,n}(\mu+\mup)\, K_n^r(\mup)\;, \label{BYBE}  
\ee 
in which we sum over the indices $k,l,m,n$.   
This equation can be illustrated by the equality of the two scattering 
diagrams in {\em figure 2}.   
 
%
%
\begin{center} 
\begin{picture}(330,210)(0,-30) 
\put(120,10){\rule{3pt}{150pt}}
\put(120,60){\line(-1,-2){25}} 
\put(120,60){\line(-1,2){50}} 
\put(120,100){\line(-2,-1){60}}
\put(120,100){\line(-2,1){60}}
\put(170,78){\line(1,0){13}} 
\put(170,82){\line(1,0){13}} 
\put(300,10){\rule{3pt}{150pt}}
\put(300,110){\line(-1,-2){50}} 
\put(300,110){\line(-1,2){25}} 
\put(300,65){\line(-2,-1){60}}
\put(300,65){\line(-2,1){60}}
\put(88,2){\shs{\fns{$A_j$}}} 
\put(52,62){\shs{\fns{$A_i$}}} 
\put(48,132){\shs{\fns{$\ol A_p$}}} 
\put(63,164){\shs{\fns{$\ol A_r$}}} 
\put(103,70){\shs{\tiny{$\ol A_k$}}} 
\put(89,98){\shs{\tiny{$\ol A_l$}}} 
\put(108,90){\shs{\tiny{$A_m$}}} 
\put(104,109){\shs{\tiny{$\ol A_n$}}} 
\put(242,2){\shs{\fns{$A_j$}}} 
\put(231,28){\shs{\fns{$A_i$}}} 
\put(226,95){\shs{\fns{$\ol A_p$}}} 
\put(270,164){\shs{\fns{$\ol A_r$}}} 
\put(266,64){\shs{\tiny{$A_k$}}} 
\put(285,53){\shs{\tiny{$A_l$}}} 
\put(288,72){\shs{\tiny{$\ol A_m$}}} 
\put(280,95){\shs{\tiny{$A_n$}}} 
\put(60,-30){\shs{\em Figure 2: The Boundary Yang--Baxter--Equation}} 
\end{picture} 
\end{center} 

There are four free indices in expression (\ref{BYBE}) and we 
thus obtain 81 independent and mostly non-trivial equations for 
the nine unknown functions $K_i^j(\mu)$.  We therefore must resort to
the use of some algebraic manipulation software in order to solve
these equations. Most of the computations in this paper were done
using MapleV. We only list the results here, details of the
derivation of these solutions can be found in appendix A.  
 
We have found essentially three different solutions to the $\ato$ 
BYBE. There is one diagonal solution and two non-diagonal solutions,
all of which contain several free coupling constant dependent
parameters and are only determined up to an overall scalar factor. 
Let us start with the two non-diagonal solutions denoted by
$K^+(\mu)$ and $K^-(\mu)$. They contain two free coupling constant
dependent parameters $g(\la)$ and $h(\la)$, and written as matrices
take the following form:   
\be 
K^{\pm}(\mu) = \left( \begin{array}{ccc}    
\kappa_{\pm}(\mu)\,\frac{h(\la)}{g(\la)} & 
e^{i\pi(\frac{\mu}3-\frac{\la}4)} &      
\pm\, e^{-i\pi(\frac{\mu}3-\frac{\la}4)}h(\la) \\ \\       
\pm\, e^{-i\pi(\frac{\mu}3-\frac{\la}4)} &   
\kappa_{\pm}(\mu)\,\frac{g(\la)}{h(\la)}     
& e^{i\pi(\frac{\mu}3-\frac{\la}4)}g(\la) \\ \\    
e^{i\pi(\frac{\mu}3-\frac{\la}4)}h(\la) &    
\pm\, e^{-i\pi(\frac{\mu}3-\frac{\la}4)}g(\la) &   
\kappa_{\pm}(\mu)h(\la)g(\la)   
\end{array}     
\right) \cA^{\pm}(\mu)\;,   \label{BYBEsol1} 
\ee    
 
\noi in which  $\cA^{\pm}(\mu)$ is an overall scalar factor, 
\[  
\kappa_{+}(\mu) = \frac{\sin(\pi(\mu-\frac{\la}4))} 
{\sin(\pi\frac{\la}2)}\;, \;\;\;\;\;\;\;\mbox{and} \;\;\;\;\;\;\; 
\kappa_{-}(\mu) = i\, \frac{\cos(\pi(\mu-\frac{\la}4))} 
{\sin(\pi\frac{\la}2)}\;.    
\label{kappa} 
\] 
The reflection matrix $\ol K_i^j(\mu)$ for the reflection of an 
incoming antisoliton into a soliton can be obtained by solving the 
following analogue of (\ref{BYBE}):   
\be  
\ol K_j^k(\mup)\, S_{\ol i,k}^{l,\ol m}(\mu+\mup)\, \ol K_m^n(\mu)\, 
S_{l,n}^{p, r}(\mu-\mup) =  
S_{\ol i,\ol j}^{\ol k,\ol l}(\mu-\mup)\, \ol K_l^m(\mu)\, S_{\ol 
k,m}^{p,\ol n}(\mu+\mup)\, \ol K_n^r(\mup)\;.  \label{BYBE2}
\ee  
This can be solved in a exactly the same way and we obtain  
 
\be    
\ol K^{\pm}(\mu) = \left( \begin{array}{ccc}    
\kappa_{\pm}(\mu)\, \frac{\ol h(\la)}{\ol g(\la)} &  
\pm\, e^{-i\pi(\frac{\mu}3-\frac{\la}4)} &     
e^{i\pi(\frac{\mu}3-\frac{\la}4)}\ol h(\la) \\ \\       
e^{i\pi(\frac{\mu}3-\frac{\la}4)} &   
\kappa_{\pm}(\mu)\, \frac{\ol g(\la)}{\ol h(\la)} &  
\pm\, e^{-i\pi(\frac{\mu}3-\frac{\la}4)}\ol g(\la) \\ \\     
\pm\, e^{-i\pi(\frac{\mu}3-\frac{\la}4)}\ol h(\la) &    
e^{i\pi(\frac{\mu}3-\frac{\la}4)}\ol g(\la) &   
\kappa_{\pm}(\mu)\, \ol h(\la) \ol g(\la)  \end{array}      
\right) \ol \cA^{\,\pm}(\mu)\;.    
\ee    
\\ 
 
Apart from these two non-diagonal solutions there is also a 
purely diagonal solution, i.e.\  $K_i^j(\mu) = \ol K_i^j(\mu) = 0$ 
(for all $i\neq j$). It can be seen easily that  in this case the BYBE 
is trivially solved by  
\be    
K^d(\mu) = \left( \begin{array}{ccc}    
1 & 0 & 0 \\ \\       
0 &  d_1(\la) & 0 \\ \\    
0 & 0 & d_2(\la) 
\end{array}     
\right) \cA^d(\mu)\;,   \label{diagsol} 
\ee    
in which $d_1(\la), d_2(\la)$ are free boundary parameters. 
Similarly, we can write a solution for the conjugate reflection 
matrix in the form  
\be    
\ol K^d(\mu) = \left( \begin{array}{ccc}    
1 & 0 & 0 \\ \\       
0 &  \ol d_1(\la) & 0 \\ \\    
0 & 0 & \ol d_2(\la) 
\end{array}     
\right) \ol \cA^d(\mu)\;,    
\ee    
 
So far no connection between the matrices $K(\mu)$ and $\ol 
K(\mu)$ exists. However, in order for these $K$-matrices to define 
consistent boundary scattering theories, several other conditions have 
to be satisfied.
We will see in the subsequent sections that these conditions will not
only provide a connection between $K(\mu)$ and $\ol K(\mu)$, but will 
also determine the overall 
scalar factors and restrict further the number of free parameters in
the reflection matrices.

\section{Boundary crossing and boundary unitarity} 

Exact $S$-matrices in two-dimensional integrable theories must 
satisfy a number of very restrictive conditions. These are known as
the analytic $S$-matrix axioms, which include the requirements of 
$S$-matrix unitarity and crossing symmetry, as well as the bootstrap 
principle. It is because of these conditions that it is possible 
to construct $S$-matrices exactly for many integrable models. These 
$S$-matrix axioms all have analogues in the case of integrable 
theories with a boundary, which impose strong restrictions on 
the possible form of the reflection matrices.  
The conditions of boundary unitarity, boundary crossing and the 
boundary bootstrap were discussed in detail by Ghoshal and 
Zamolodchikov in \cite{ghosh94}. 
In this section we will use these conditions in order to construct 
consistent reflection matrices related to the above solutions of the
BYBE.    

\subsection{Boundary unitarity} 

The condition of boundary unitarity provides a connection between 
$K(\mu)$ and $\ol K(\mu)$, and takes the form 
\be 
K_i^j(\mu)\ol K_j^{\,k}(-\mu) = \de_i^k\;. \label{bunit} 
\ee   
Putting our solutions from the previous section into this equation we 
first find that it can only be satisfied if $K^+(\mu)$, $K^-(\mu)$ and 
$K^d(\mu)$ are connected with $\ol K^{\,+}(\mu)$, $\ol K^{\,-}(\mu)$
and $\ol K^{\,d}(\mu)$, respectively. 
Furthermore, it emerges that (\ref{bunit}) also puts a constraint on the 
free boundary parameters, namely for the non-diagonal solutions we find 
\be 
g(\la)\ol g(\la) = h(\la)\ol h(\la) = 1\;, 
\ee 
and in the case of $K^d(\mu)$  
\be  
d_1(\la)\, \ol d_1(\la) = d_2(\la)\, \ol d_2(\la) = 1\;.  
\ee 
 
\noi And finally, the conditions imposed on the overall scalar factors 
by (\ref{bunit}) are 
\bea 
\cA^+(\mu)\ol \cA^{\,+}(-\mu) &=& - \frac{\sin^2\(\pi\frac{\la}2\)} 
{\sin\(\pi(\mu+\frac34 \la)\) \sin\(\pi(\mu-\frac34 \la)\)}\;, 
\nn \\ 
\cA^-(\mu)\ol \cA^{\,-}(-\mu) &=& - \frac{\sin^2\(\pi\frac{\la}2\)} 
{\cos\(\pi(\mu+\frac34 \la)\) \cos\(\pi(\mu-\frac34 \la)\)}\;, \nn \\   
\cA^d(\mu)\, \ol \cA^d(-\mu) &=& 1\;. \label{sfact1}  
\eea

\subsection{Boundary crossing} 

The boundary analogue of the crossing symmetry condition was called 
boundary cross-unitarity by Ghoshal and Zamolodchikov. In 
our notation this condition takes the form 
\bea 
K_i^j(\frac34 \la -\mu) = S_{j,i}^{k,l}(2\mu)K_l^k(\frac34 
\la+\mu)\;, \nn \\ \nn \\ 
\ol K_i^j(\frac34 \la -\mu) = S_{\ol j\,,\ol i}^{\ol k\,,\ol 
l}(2\mu)\ol K_l^k(\frac34 \la+\mu)\;. \label{bcross} 
\eea 
Putting the solutions to the BYBE and the explicit expression for the 
$S$-matrix, as given  
in section 2, into these equations, we obtain a second set of 
equations determining the scalar factors:  
\bea 
\frac{\cA^+(-\mu+\frac34 \la)}{\cA^+(\mu+\frac34 \la)} &=& \frac{\ol    
\cA^{\,+}(-\mu+\frac34 \la)}{\ol \cA^{\,+}(\mu+\frac34 \la)} =  - 
\frac{\sin(\pi(\mu+\frac{\la}2))} {\sin(\pi(\mu-\frac{\la}2))}\,   
F(2\mu)\;, \nn \\ \nn \\ 
\frac{\cA^-(-\mu+\frac34 \la)}{\cA^-(\mu+\frac34 \la)} &=& \frac{\ol    
\cA^{\,-}(-\mu+\frac34 \la)}{\ol \cA^{\,-}(\mu+\frac34 \la)} = 
\frac{\cos(\pi(\mu+\frac{\la}2))} {\cos(\pi(\mu-\frac{\la}2))}\, 
F(2\mu)\;, \nn \\ \nn \\ 
\frac{\cA^d(-\mu+\frac 34 \la)} {\cA^d(\mu+\frac 34 \la)} &=& 
\frac{\ol \cA^d(-\mu+\frac 34 \la)} {\ol \cA^d(\mu+\frac 34 \la)} = 
F(2\mu)\;, \label{sfact2}   
\eea 
in which $F(\mu)$ is the $S$-matrix scalar factor (\ref{F11}).  
Here we already notice one major difference between this case and the 
case of the boundary sine--Gordon theory \cite{ghosh94}, in that none 
of the equations in (\ref{sfact1}, \ref{sfact2}) contain any
dependence on the free boundary parameters $g(\la)$, $h(\la)$ or
$d_1(\la)$, $d_2(\la)$.

\subsection{The overall scalar factors} 

We want to construct solutions to equations (\ref{sfact1}) and
(\ref{sfact2}). In order to do this it will prove useful to separate
these two sets of equations by making the following ansatz:   
\bea  
\cA^{\ep}(\mu) &=& \sin(\pi\frac{\la}2)a^{\ep}_0(\mu)a_1(\mu)\;, \nn 
\\    
\ol \cA^{\ep}(\mu) &=& \sin(\pi\frac{\la}2)\ol a^{\,\ep}_0(\mu)\ol 
a_1(\mu)\;,  
\label{separate}  
\eea  
in which $\ep = +,-$ or $d$, such that $a_1(\mu)$, which is the same 
for all three cases, is determined by  
\bea  
a_1(\mu)\ol a_1(-\mu) &=& 1 \;, \nn \\ \nn \\  
\frac{a_1(-\mu+\frac34 \la)}{a_1(\mu+\frac34 \la)} &=&  
\frac{\ol a_1(-\mu+\frac34 \la)}{\ol a_1(\mu+\frac34 \la)} =  
- F(2\mu)\;,    
\eea  
and the $a_0^{\ep}(\mu)$ satisfy 
\bea  
a^+_0(\mu)\ol a^{\,+}_0(-\mu) &=& \frac{-1}{\sin(\pi(\mu+\frac34\la))   
\sin(\pi(\mu-\frac34\la))}\;, \nn \\ \nn \\ 
\frac{a^+_0(-\mu+\frac34 \la)}{a^+_0(\mu+\frac34 \la)} &=& \frac{\ol    
a^{\,+}_0(-\mu+\frac34 \la)}{\ol a^{\,+}_0(\mu+\frac34 \la)} =   
\frac{\sin(\pi(\mu+\frac{\la}2))} {\sin(\pi(\mu-\frac{\la}2))}\;, 
\label{a0p}  
\eea  
and analogously 
\bea 
a^-_0(\mu)\ol a^{\,-}_0(-\mu) &=& \frac{-1}{\cos(\pi(\mu+\frac34\la))   
\cos(\pi(\mu-\frac34\la))}\;, \nn \\ \nn \\ 
\frac{a^-_0(-\mu+\frac34 \la)}{a^-_0(\mu+\frac34 \la)} &=& \frac{\ol    
a^{\,-}_0(-\mu+\frac34 \la)}{\ol a^{\,-}_0(\mu+\frac34 \la)} =  - 
\frac{\cos(\pi(\mu+\frac{\la}2))} {\cos(\pi(\mu-\frac{\la}2))}\;, 
\label{a0m} \\ \nn \\  
a^d_0(\mu)\ol a^{\,d}_0(-\mu) &=& \frac{1}{\sin^2(\pi\frac{\la}2)}\;,
\nn \\ \nn \\  
\frac{a^d_0(-\mu+\frac34 \la)}{a^d_0(\mu+\frac34 \la)} &=& \frac{\ol    
a^{\,d}_0(-\mu+\frac34 \la)}{\ol a^{\,d}_0(\mu+\frac34 \la)} =  -1\;. 
\label{a0d}  
\eea  
\vs{5pt}
 
\noi Solutions to equations of this type can be constructed in a
similar way as the solutions for the scalar factors in imaginary ATFTs
on the whole line. Let us concentrate on $a_0^{+}(\mu)$ first. It is
easy to check that equations (\ref{a0p}) can be solved formally by an
expression of the form 
\[   
a^+_0(\mu) =\ol a^{\,+}_0(\mu) = \frac1{\sin(\pi(\mu-\frac{\la}4)) 
\sin(\pi(\mu-\frac34 {\la})) }  
\prod_{j=1}^{\infty} \frac{f(\mu+3j\la-3\la)  
f(-\mu+3j\la-\frac32\la)} {f(-\mu+3j\la)  
f(\mu+3j\la-\frac32\la)}\;,     
\]  
for any function $f(\mu)$ with   
\[  
f(\mu)f(-\mu) = \sin(\pi(\mu-\frac{\la}4))  
\sin(\pi(-\mu-\frac{\la}4))\;. 
\]   
In order to avoid the occurrence of an infinite number of poles in the 
physical strip, the obvious choice for $f(\mu)$ is 
\[ 
f(\mu) = \frac{\pi} 
{\Ga(\mu-\frac{\la}4) \Ga(1+\mu+\frac{\la}4)}\;.  
\] 
 
\noi Using the expression (\ref{F11}) for the $S$-matrix scalar factor
the solution for $a_1(\mu)$ can be constructed in a similar way and we
obtain   
\bea  
a_1(\mu) = \ol a_1(\mu) &=& \prod_{k=1}^{\infty}  
\frac{\Ga(2\mu+6k\la-\frac92 \la+1)   
\Ga(2\mu+6k\la-\frac32 \la)} {\Ga(-2\mu+6k\la-\frac92 \la+1)  
\Ga(-2\mu+6k\la-\frac32 \la)} \nn \\  
&& \hs{8pt}\times \frac{\Ga(-2\mu+6k\la-\frac72 \la+1) 
\Ga(-2\mu+6k\la-\frac52 \la)}  {\Ga(2\mu+6k\la-\frac72 \la+1) 
\Ga(2\mu+6k\la-\frac52 \la)} \;. \label{a1}  
\eea  
If we use the abbreviation 
\be 
G_j(\mu,a) \equiv \frac{\Ga(\mu+3j\la+a)}{\Ga(-\mu+3j\la+a)}\;, 
\ee  
the final result for the scalar factor can be written in the following 
compact form: 
\bea 
\cA^+(\mu) = \ol{\cA}^{\,+} (\mu) &=& \frac{\sin(\pi\frac{\la}2)} 
{\sin(\pi(\mu-\frac34 {\la}))}\,   
\prod_{j=1}^{\infty} \frac{G_j(\mu,-\frac 74 \la)\, G_j(\mu,-\frac 
54 \la+1)} {G_j(\mu,-\frac{\la}4)\, G_j(\mu,-\frac{11}4 \la+1)} \nn \\  
&& \hs{70pt} \times \prod_{k=1}^{\infty} \frac{G_{2k}(2\mu,-\frac 32
\la)\,  
G_{2k}(2\mu,-\frac 92 \la+1)} {G_{2k}(2\mu,-\frac 52 \la)\, 
G_{2k}(2\mu,-\frac 72 \la+1)}\;. \label{Ap} 
\eea 
 
\noi Analogously, for the case of $K^-(\mu)$ we find 
\bea 
\cA^-(\mu) = - \ol{\cA}^{\,-} (\mu) &=& \frac{\sin(\pi\frac{\la}2)} 
{\cos(\pi(\mu-\frac34 {\la}))}\, \bl - \frac 34 \la \br\,   
\prod_{j=1}^{\infty} \frac{G_j(\mu,-\frac 74 \la+\frac 12)\, 
G_j(\mu,-\frac 54 \la+\frac 12)} {G_j(\mu,-\frac{\la}4+\frac 12)\, 
G_j(\mu,-\frac{11}4 \la+\frac 12)} \nn \\   
&& \hs{112pt} \times \prod_{k=1}^{\infty} \frac{G_{2k}(2\mu,-\frac 32
\la)\,  
G_{2k}(2\mu,-\frac 92 \la+1)} {G_{2k}(2\mu,-\frac 52 \la)\, 
G_{2k}(2\mu,-\frac 72 \la+1)}\;,\hs{30pt}  \label{A1m} 
\eea 

\noi in which we have used the bracket notation (\ref{bracket1}) as
introduced in section 2.  
The solution for $a_0^d(\mu)$ is rather trivial and we obtain the
scalar factor for the diagonal reflection matrices in the form
\be 
\cA^d(\mu) = \ol \cA^d(\mu) = \bl - \frac 34 \la\br\, a_1(\mu)\;. 
\ee 
 
As usual, the infinite products of Gamma functions in these overall 
scalar factors can alternatively be written in integral form. We find   
\bea 
\cA^+(\mu) &=& \frac{\sin(\pi\frac{\la}2)} {\sin(\pi(\mu-\frac34 \la))} 
\exp  \int_0^{\infty} \frac{dt}t \left[-2\sinh(2\mu t){\cal 
I}_1(t)+ \sinh(\mu t) {\cal I}_+(t)\right] \;, \nn \\ \nn 
\\  
\cA^-(\mu) &=& \frac{\sin(\pi\frac{\la}2)} {\cos(\pi(\mu-\frac34 
\la))}\, \bl -\frac 34 \la \br\, 
\exp \int_0^{\infty} \frac{dt}t \left[-2\sinh(2\mu t) {\cal 
I}_1(t)+ \sinh(\mu t) {\cal I}_-(t)\right] \;, \nn \\ \nn 
\\  
\cA^d(\mu) &=& \bl -\frac 34 \la \br\, 
\exp  \int_0^{\infty} \frac{dt}t \left[-2\sinh(2\mu t) {\cal I}_1(t) 
\right] \;,   
\eea 
 
in which  
\bea 
{\cal I}_1(t) &=& \frac{\sinh(\frac{\la}2t) \sinh((\la-\frac 12) 
t)}{\sinh(\frac 12 t) 
\sinh(3\la t)} \;, \nn \\ 
{\cal I}_+(t) &=& \frac{\sinh(\frac{\la-1}2 t)}{\sinh(\frac 12 t) 
\cosh(\frac 34 \la t)} \;, \nn \\ 
{\cal I}_-(t) &=& \frac{\sinh(\frac{\la}2t)} {\sinh(\frac 12 t) 
\cosh(\frac 43 \la t)} \;. 
\eea

Of course, in all three cases there still exists an overall
ambiguity. We can always multiply $\cA(\mu)$ and $\ol{\cA}(\mu)$ by a
function $\si(\mu)$ which satisfies  
\bea 
\si(\mu)\si(-\mu) &=& 1\;, \nn \\ 
\si(\mu+\frac 34\la)&=& \si(-\mu+\frac 34 \la)\;. \label{sigma1} 
\eea 
In the next section the bootstrap condition is used to remove this 
ambiguity.  
Note also that all solutions we found satisfy $\cA(\mu) = \pm \ol 
\cA(\mu)$. There is however no obvious reason that this has to be so, 
and we will see below that there are more general scalar factors for 
which $\cA(\mu) \neq \pm \ol \cA(\mu)$.

\section{The boundary bootstrap relations}  

The bootstrap principle for an integrable two-dimensional field
theory states that simple \mbox{$S$-matrix} poles in the physical
strip ($0 
\leq  \mbox{Im}(\th) \leq \pi$) correspond to bound states of the two 
incoming particles. In the bulk theory     
the bootstrap equations relate the scattering amplitude of such a 
bound state to the product of the scattering amplitudes of the 
incoming states, thus allowing the construction of bound state 
$S$-matrices from the $S$-matrices of the fundamental states in the 
theory.    
Correspondingly, in theories on a half-line the so-called boundary 
bootstrap equations can be used to construct the amplitudes 
for the reflection of bound states from the boundary.    
 
As mentioned earlier, the \uq{\ato} invariant scattering theory 
contains two types of bound states, namely scalar bound states and 
bound states with non-zero topological charge, the so-called excited 
solitons. Therefore, we have to examine two types of bootstrap 
equations, the soliton bootstrap and the breather bootstrap.

\subsection{The soliton bootstrap}  

Again, here we are only interested in the bound states with lowest 
mass. The lowest mass excited solitons are just the fundamental  
solitons themselves. In other words in the \uq{\ato} invariant 
scattering  
theory two solitons can fuse into a single antisoliton and vice versa.  
This fusion corresponds to the physical strip pole at $\mu =\la$
 (or $\th=i\frac{2\pi}3$) in $S^T(\mu)$. In \cite{gande95}   
it was shown that this fusion can formally be expressed in terms of 
the FZ generators:
\bea  
\ol A_j(\frac{\mu_1+\mu_2}2) &=& \lim_{\mu_2-\mu_1 \lra \la}  
\left[A_k(\mu_1)A_l(\mu_2)+\ga_{k,l} A_l(\mu_1)A_k(\mu_2)\right]\;,  
\nn \\  
A_j(\frac{\mu_1+\mu_2}2) &=& \lim_{\mu_2-\mu_1 \lra \la}  
\left[\ol A_l(\mu_1)\ol A_k(\mu_2)+\ga_{k,l} \ol A_k(\mu_1)\ol  
A_l(\mu_2)\right]\;,   
\label{solfuse}   
\eea  
in which $(j,k,l)$ are even permutations of $(1,2,3)$ and  
$\ga_{k,l} = e^{i\pi\frac{\la}3\,\nu(k,l)\,}$, where $\;\nu(k,l)\,$ was 
defined in (\ref{nukl}).      
  
By analogy with the bootstrap on the whole line, the boundary
bootstrap expresses the fact that the reflection amplitude of two
particles should be independent of whether their fusion into a bound
state occurs before or after their reflection from the boundary. This
can be illustrated in the equality of the two reflection processes
depicted in {\em figure 3}.   
The corresponding soliton bootstrap equations provide a relation 
between $K(\mu)$ and $\ol K(\mu)$ and can therefore not only serve as 
a highly non-trivial check for our reflection matrices, but they also 
fix the CDD ambiguity (\ref{sigma1}) in the overall scalar factors.  
This will be important in the next section, since different choices of 
$\si(\mu)$ would lead to different breather   
reflection amplitudes.  
  
%
%
\begin{center} 
\begin{picture}(330,190)(0,-30) 
\put(120,10){\rule{3pt}{140pt}}
\put(120,80){\line(-1,-1){30}} 
\put(90,50){\line(-2,-1){60}} 
\put(90,50){\line(-1,-2){15}} 
\put(120,80){\line(-1,1){30}} 
\put(90,110){\line(-2,1){60}} 
\put(90,110){\line(-1,2){15}} 
\put(160,78){\line(1,0){13}} 
\put(160,82){\line(1,0){13}} 
\put(300,10){\rule{3pt}{140pt}}
\put(300,80){\line(-4,-3){80}} 
\put(300,80){\line(-4,3){24}} 
\put(300,50){\line(-1,-2){15}} 
\put(300,50){\line(-1,2){24}} 
\put(276,98){\line(-1,1){26}} 
\put(250,124){\line(-2,1){35}} 
\put(250,124){\line(-1,2){10}} 
\put(22,9){\shs{\fns{$A_j$}}} 
\put(69,9){\shs{\fns{$A_k$}}} 
\put(212,9){\shs{\fns{$A_j$}}} 
\put(280,9){\shs{\fns{$A_k$}}} 
\put(97,68){\shs{\scs{$\ol Ai$}}} 
\put(106,96){\shs{\scs{$A_l$}}} 
\put(264,111){\shs{\scs{$A_l$}}} 
\put(268,83){\shs{\scs{$\ol A_m$}}} 
\put(286,92){\shs{\scs{$\ol A_n$}}} 
\put(285,55){\shs{\scs{$\ol A_p$}}} 
\put(287,78){\shs{\tiny{$A_r$}}} 
\put(22,145){\shs{\fns{$\ol A_m$}}} 
\put(69,145){\shs{\fns{$\ol A_n$}}} 
\put(204,145){\shs{\fns{$\ol A_m$}}} 
\put(238,145){\shs{\fns{$\ol A_n$}}} 
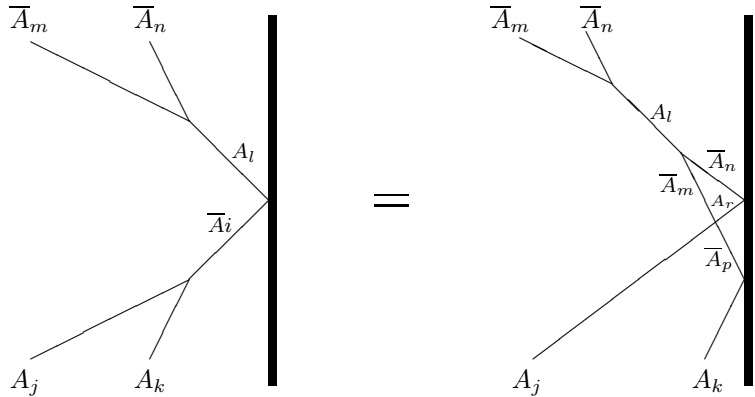
\put(60,-30){\shs{\em Figure 3: The boundary soliton bootstrap
equations}} 
\end{picture} 
\end{center} 

The evaluation of the bootstrap relations involves mainly  
straightforward but rather tedious computations. 
We therefore only provide the main results here, and more details can
be found in appendix B.    
First we find that the soliton bootstrap equations imply that in the  
case of $K^+(\mu)$ we have to include a factor $\si(\mu)$ into 
the overall scalar factor. This additional factor must satisfy 
equations (\ref{sigma1}) and   
\be  
\si(\mu+\frac{\la}2)\si(\mu-\frac{\la}2) =  \bl -\frac{\la}4 \br^2 \bl  
\frac34 \la \br^2 \bl -\frac54 \la \br^2\, \si(\mu)\;. \label{sigma2}  
\ee  
A simple solution to these equations can be written
conveniently in the form
\be  
\si^+(\mu) =  \bl \frac{\la}4 \br \bl \frac54 \la
\br\;. \label{sigmap}  
\ee  
For the cases of $K^-(\mu)$ and $K^d(\mu)$ the soliton bootstrap
relations are satisfied without introducing any further factors. 
Furthermore, we find that the soliton bootstrap equations put another 
constraint on the free parameters. We obtain that these bootstrap 
equations can only be satisfied if we choose  
\bea  
g(\la) &=& -\frac 1{h(\la)}\;, \hs{40pt} (\mbox{in} \;\; K^+)\;, \nn \\   
g(\la) &=& \frac i{h(\la)}\;, \hs{50pt} (\mbox{in} \;\; K^-)\;, \nn \\ 
d_1(\la) &=& \frac 1{d_2(\la)}\;, \hs{48pt} (\mbox{in} \;\; K^d)\;,
\label{solrestr}  
\eea  
which reduces the total number of continuous free parameters in each 
of the reflection matrices to one.  
We, therefore, obtain the final result for the \uq{\ato} invariant  
reflection matrices:\\
   
\bea     
K^+(\mu) &=& \left( \begin{array}{ccc}     
- \frac{\sin(\pi(\mu-\frac{\la}4))}{\sin(\pi\frac{\la}2)}\, h^2(\la) &  
e^{i\pi(\frac{\mu}3-\frac{\la}4)} &       
e^{-i\pi(\frac{\mu}3-\frac{\la}4)}h(\la) \\ \\        
e^{-i\pi(\frac{\mu}3-\frac{\la}4)} &    
- \frac{\sin(\pi(\mu-\frac{\la}4))}{\sin(\pi\frac{\la}2)}\,  
h^{-2}(\la)  & -e^{i\pi(\frac{\mu}3-\frac{\la}4)}\, h^{-1}(\la) \\ \\      
e^{i\pi(\frac{\mu}3-\frac{\la}4)}h(\la) &     
-e^{-i\pi(\frac{\mu}3-\frac{\la}4)}\, h^{-1}(\la) &    
- \frac{\sin(\pi(\mu-\frac{\la}4))}{\sin(\pi\frac{\la}2)}  
\end{array}      
\right) \si^+(\mu)\, \cA^+(\mu)\;, \nn \\ \label{Kp} \\ \nn \\ \nn \\
K^-(\mu) &=& \left( \begin{array}{ccc}     
\frac{\cos(\pi(\mu-\frac{\la}4))}{\sin(\pi\frac{\la}2)} \,h^2(\la) &  
e^{i\pi(\frac{\mu}3-\frac{\la}4)} &       
-e^{-i\pi(\frac{\mu}3-\frac{\la}4)}h(\la) \\ \\        
-e^{-i\pi(\frac{\mu}3-\frac{\la}4)} &    
-\frac{\cos(\pi(\mu-\frac{\la}4))}{\sin(\pi\frac{\la}2)}  
\, h^{-2}(\la)       
& ie^{i\pi(\frac{\mu}3-\frac{\la}4)} h^{-1}(\la) \\ \\     
e^{i\pi(\frac{\mu}3-\frac{\la}4)}h(\la) &     
-ie^{-i\pi(\frac{\mu}3-\frac{\la}4)}\, h^{-1}(\la) &    
-\frac{\cos(\pi(\mu-\frac{\la}4))}{\sin(\pi\frac{\la}2)}  
\end{array}      
\right) \cA^-(\mu)\;, \nn \\ \label{Km}    
\eea          
The corresponding reflection matrices  $\ol K^{\,\pm}(\mu)$ for the 
charge conjugate states can be written conveniently as 
\be  
\ol K^{\,\pm}(\mu) = {\cal H}^{\pm}\,\(K^{\pm}(\mu)\)^{\top}\,{\cal  
H}^{\pm}\;,    
\ee  
in which   
\[  
{\cal H}^+ = \(\begin{array}{ccc} h^{-2}(\la) & 0 & 0 \\ 0 &  
h^2(\la) & 0 \\ 0 & 0 & 1 \end{array}\)\;, \;\;\;\;\;\; \mbox{and}  
\;\;\;\;\; {\cal H}^- = \(\begin{array}{ccc} ih^{-2}(\la) & 0 & 0 \\ 0  
& -ih^2(\la) & 0 \\ 0 & 0 & -i \end{array}\)\;.  
\]  
  
\noi The diagonal reflection matrices are  
\be    
K_d(\mu)= \left( \begin{array}{ccc}    
1 & 0 & 0 \\ \\       
0 &  d(\la) & 0 \\ \\    
0 & 0 & \frac 1{d(\la)} 
\end{array}     
\right) \cA^d(\mu)\;, \;\;\;\;\;\;\;\;\;\;   
\ol K^d(\mu) = \left( \begin{array}{ccc}    
1 & 0 & 0 \\ \\       
0 &  \frac 1{d(\la)} & 0 \\ \\    
0 & 0 & d(\la) 
\end{array}     
\right) \cA^d(\mu)\;.    
\ee 
  
However, as already mentioned in the previous section, there is
another ambiguity in the overall scalar factors which cannot be
removed by the requirements of the bootstrap equations. The above
scalar factors all satisfy $\cA(\mu) = \pm \ol \cA(\mu)$.  
Although this seems to be a reasonable assumption, it does not follow 
from the equations (\ref{sfact1}) and (\ref{sfact2}). 
We are thus free to define more general scalar factors  
\bea 
\cA_n^{\ep}(\mu) &\equiv& \tau(\mu) \cA^{\ep}(\mu)\;, \nn \\ 
\ol \cA_n^{\,\ep}(\mu) &\equiv& \ol \tau(\mu) \ol \cA^{\,\ep}(\mu)\;,
\label{Anew} 
\eea 
in which $\ep$ can be $+,-$ or $d$. These new scalar factors
satisfy the neccesary unitarity and crossing conditions if we require
that  
\bea 
\tau(\mu) \ol\tau(-\mu) &=& 1\;, \label{tau1} \\ 
\frac{\tau(-\mu+\frac 34\la)}{\tau(\mu+\frac 34\la)} &=& 
\frac{\ol \tau(-\mu+\frac 34\la)}{\ol \tau(\mu+\frac 34\la)} = 1\;, 
\label{tau2}  
\eea 
However, we also need to ensure that they do not
violate the soliton bootstrap relations. From the detailed computation
in appendix B we can see that the soliton bootstrap equations will
continue to be satisfied if we require
\bea 
\tau(\mu-\frac{\la}2)\tau(\mu+\frac{\la}2) &=& \ol\tau(\mu)\;, \nn \\ 
\ol\tau(\mu-\frac{\la}2)\ol\tau(\mu+\frac{\la}2) &=& 
\tau(\mu)\;. \label{tau3}  
\eea 
The question remains whether any solutions to  
equations (\ref{tau1} -- \ref{tau3}) exist. It is straightforward to
see that there are in fact infinitely many such solutions which can be
written in the following form:   
\be 
\tau(\mu) = \frac{\lb \eta \rb \lb -\eta+\frac 32\la\rb} {\lb \eta
+\la \rb \lb -\eta+ \frac{\la}2 \rb}\;,  \label{tausol}   
\ee 
in which $\lb a \rb \equiv \sin(\frac{\pi}{3\la}(\mu+a))$ and   
$\eta = \eta(\la)$ is an arbitrary function of $\la$.  
Equation (\ref{tau1}) then implies  
\be 
\ol\tau(\mu) = \frac{\lb -\eta-\la \rb \lb \eta-\frac{\la}2 \rb} {\lb  
-\eta \rb \lb \eta-\frac 32\la\rb} \;,      
\ee 
and is straightforward to check that this satisfies equations 
(\ref{tau2}) and (\ref{tau3}). We will discuss the implications of
this ambiguity in the next section.

\subsection{The breather bootstrap}   

The breather bootstrap equations are similar to the soliton bootstrap,  
but now the two incoming solitons fuse into a scalar bound state, the 
so-called breather, instead of another soliton. Again we only consider 
the lowest mass breathers whose FZ generators where defined in 
(\ref{FZB}).  
  
The breather bootstrap equations on the boundary are depicted in {\em
figure 4}. It is important to note here that the type of breather
created depends on the ordering of the incoming soliton--antisoliton
pair, which can also be seen from the definition (\ref{FZB}).    
The picture in {\em figure 4} now illustrates clearly the fact that
the change of multiplets in the soliton reflection implies that the 
breather reflection is purely diagonal, i.e.\ a breather $B_1$ can 
only reflect into itself and not into its conjugate partner $\ol B_1$. 
 
%
%
\begin{center} 
\begin{picture}(330,190)(0,-30) 
\put(120,10){\rule{3pt}{140pt}}
\put(120,80){\line(-1,-1){30}} 
\put(90,50){\line(-2,-1){60}} 
\put(90,50){\line(-1,-2){15}} 
\put(120,80){\line(-1,1){30}} 
\put(90,110){\line(-2,1){60}} 
\put(90,110){\line(-1,2){15}} 
\put(160,78){\line(1,0){13}} 
\put(160,82){\line(1,0){13}} 
\put(300,10){\rule{3pt}{140pt}}
\put(300,80){\line(-4,-3){80}} 
\put(300,80){\line(-4,3){24}} 
\put(300,50){\line(-1,-2){15}} 
\put(300,50){\line(-1,2){24}} 
\put(276,98){\line(-1,1){26}} 
\put(250,124){\line(-2,1){35}} 
\put(250,124){\line(-1,2){10}} 
\put(22,9){\shs{\fns{$A_m$}}} 
\put(69,9){\shs{\fns{$\ol A_m$}}} 
\put(212,9){\shs{\fns{$A_m$}}} 
\put(280,9){\shs{\fns{$\ol A_m$}}} 
\put(96,69){\shs{\scs{$B_1$}}} 
\put(105,97){\shs{\scs{$B_1$}}} 
\put(266,109){\shs{\scs{$B_1$}}} 
\put(285,57){\shs{\tiny{$A_j$}}} 
\put(292,68){\shs{\tiny{$\ol A_l$}}} 
\put(273,80){\shs{\tiny{$A_k$}}} 
\put(287,92){\shs{\tiny{$\ol A_k$}}} 
\put(22,145){\shs{\fns{$\ol A_m$}}} 
\put(69,145){\shs{\fns{$A_m$}}} 
\put(204,147){\shs{\fns{$\ol A_m$}}} 
\put(238,147){\shs{\fns{$A_m$}}} 
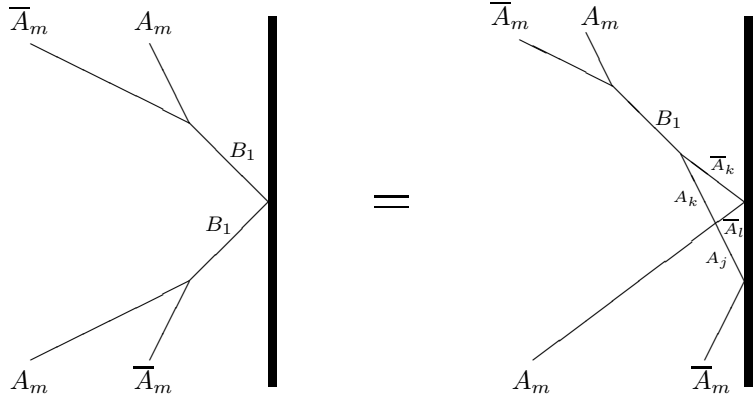
\put(50,-30){\shs{\em Figure 4: The boundary breather bootstrap
equations}} 
\end{picture} 
\end{center} 
%
%
  

Using the FZ algebra relation it is now straightforward to evaluate 
the two scattering processes in {\em figure 4}, and we obtain the 
reflection amplitudes for the reflection of a breather $B_1$ from the 
boundary.  
The details of these computations can again be found in appendix B, 
where we obtained the following results 
\be  
B_1(\mu) \cB = K^{(\ep)}_B(\mu) B_1(-\mu) \cB\;,  
\ee  
in which the boundary reflection factors corresponding to the three 
different reflection matrices are given by 
\bea  
K^{(+)}_B(\mu) &=& \bl \frac 12\br\, \bl -\la \br\,   
\bl \la - \frac 12\br\;,  \label{KBp} \\ \nn \\ 
K^{(-)}_B(\mu) &=& - \bl \frac{\la}2\br\, \bl -\frac{\la}2-\frac 12 \br\,   
\bl -\frac 32 \la + \frac 12\br\;, \label{KBm}  \\ \nn \\  
K^{(d)}_B(\mu) &=& \bl -\frac{\la}2 - \frac 12 \br\, \bl -\la \br\,   
\bl -\frac 32 \la + \frac 12\br\;. \label{KBd} 
\eea  
Here we have again used the block notation as defined in 
(\ref{bracket1}).   
  
Finally, let us discuss briefly how the inclusion of the more general 
scalar factors (\ref{Anew}) would  
affect the final result of the breather reflection factors. 
From the breather bootstrap equations 
we find that using the new scalar factors $\cA_n(\mu)$, $\ol 
\cA_n(\mu)$ introduces an additional term 
\be 
k^{(\eta)}(\mu) \equiv \tau(\mu-\frac 34 \la+\frac 12) 
\ol\tau(\mu+\frac 34\la-\frac 12) =  
\bl \eta-\frac 34 \la+\frac 12\br\, \bl -\eta+\frac 34 \la+\frac 12\br\, 
\bl -\eta-\frac{\la}4 -\frac 12\br\, \bl \eta+\frac{\la}4 -\frac
12\br  \label{keta}
\ee 
into the lowest breather reflection factor $K^{(\ep)}_B(\mu)$. 
Note that we can of course use any number of terms of the form 
(\ref{tausol}) as our $\tau(\mu)$, which would then introduce more 
than one free parameter into the breather reflection factors. 
However the inclusion of these more general scalar factors also
introduces new poles into the reflection factors and it 
would therefore be necessary to investigate whether these  
additional physical strip poles can be explained in terms of physical
scattering processes.

\section{Real $a_2^{(1)}$ affine Toda field theory on a half-line} 

Affine Toda field theory with real coupling constant based on the 
affine Lie algebra $\ato$ is an integrable two-dimensional field
theory which contains two (mass  degenerate) scalar particles. 
Extending earlier works by Arinshtein et al.\ \cite{arins79},
exact $S$-matrices for this theory (and all other ATFTs based on
simply laced algebras) have been constructed  
some years ago by Braden et al.\ in \cite{brade90}. Some years later
it was demonstrated in \cite{gande95} that these $S$-matrices could
also be obtained from the \uq{\ato} invariant \mbox{$S$-matrix}.
It was shown that the scattering amplitudes of the lowest breathers
were identical to the real affine Toda $S$-matrices after analytic
continuation of the coupling constant from imaginary to real value.    
 
In this section this identification of the breathers with the 
fundamental quantum particles will be used in order to provide a 
conjecture for the reflection factors in real $\ato$ ATFT.  
We therefore have to analytically continue the above breather 
reflection factors (\ref{KBp} -- \ref{KBd}) from imaginary to real 
coupling constant $\beta$. The connection between $\la$ 
and the Toda coupling constant $\beta$ must be the same as in the bulk
theory and was given in (\ref{labeta}). 
In order to compare our results with previous works on real ATFT we  
use instead of (\ref{bracket1}) a slightly different form of the
bracket notation in this section  
\be 
\bl y \br \equiv \frac{\sin(\frac{\th}{2i}+\frac{y\pi}6)} 
{\sin(\frac{\th}{2i}-\frac{y\pi}6)}\;. 
\ee   
 
The coupling constant dependence in the real Toda $S$-matrices is 
usually given in terms of the function 
\be 
B(\beta) = \frac1{2\pi}\frac{\beta^2}{1+\frac{\beta^2}{4\pi}}\;, 
\ee 
and thus analytic continuation $\beta \lra i\beta\;$ is
equivalent to   
\be 
\la \lra -\frac{2}{B}\;. 
\ee 
Performing this analytic continuation in the three different 
breather reflection factors from the previous section, we obtain three
possible reflection factors   
\bea 
K^{(+)}_B(\mu) \hs{10pt} \lra \hs{10pt} K^{(+)}_1(\th) &=& \bl -2\br\, 
\bl -\frac B2\br\, \bl 2+\frac B2\br \;, \nn \\ \nn \\ 
K^{(-)}_B(\mu)  \hs{10pt} \lra \hs{10pt} K^{(-)}_1(\th) &=& - \bl 
1\br\, \bl \frac B2-1\br\, \bl 3- \frac B2\br \;, \nn \\ \nn \\ 
K^{(d)}_B(\mu)  \hs{10pt} \lra \hs{10pt} K^{(d)}_1(\th) &=& \bl 
-2\br\, \bl \frac B2-1\br\, \bl 3-\frac B2\br \;, \label{realref1} 
\eea 
in which the $K_1$'s are now the reflection amplitudes for the
reflection of the first particle in real $\ato$ ATFT on a half-line. 

As an important check for these reflection factors we need to 
demonstrate their consistency with the boundary bootstrap.  
In $a_2^{(1)}$ ATFT we have possible fusion processes $1+1 \lra 2$ and
$2+2\lra 1$, occurring at the rapidity  
difference $\th = i\frac{2\pi}3$, and we thus require the following 
reflection bootstrap relation: 
\be 
K_2(\th) = K_1(\th+\frac{i\pi}3)\, K_1(\th-\frac{i\pi}3)\, 
S_{11}(2\th)\;, 
\ee 
in which $S_{11}(\th)$ is the $\ato$ affine Toda $S$-matrix
\cite{brade90}  
\be 
S_{11}(\th) = \bl 2 \br\, \bl -2+B\br\, \bl -B \br\;, 
\ee 
and therefore 
\[ 
S_{11}(2\th) =  - \bl 1\br\, \bl -2\br\, \bl -1+\frac B2  \br\, \bl
2+\frac B2  \br\, \bl -\frac B2 \br\, \bl 3-\frac B2\br\;.   
\] 
Using this expression and the above formulas for $K_1(\th)$ we find in
all three cases  
\bea 
K^{(\ep)}_2(\th) &=& K^{(\ep)}_1(\th)\;, \label{K1=K2}  
\eea 
which is expected since the two particles in $\ato$ ATFT are mass 
degenerate and the boundary
conditions do not distinguish between the two particles\footnote 
{Since the second quantum particle is identified with the conjugate 
breather $\ol B_1$, we could have checked relation (\ref{K1=K2}) 
directly by performing the breather bootstrap on $\ol B_1$, and we 
would have found that $K_B(\mu) = K_{\ol B}(\mu)$.}.  
Furthermore, the requirement of boundary crossing and unitarity 
implies  
\be 
K_1^{(\ep)}(\th)\,K_2^{(\ep)}(\th-i\pi) =  S_{11}(2\th)\;,  
\ee 
which again can be checked to be true for all three reflection factors 
in (\ref{realref1}). 
 
As already mentioned in section 2, one of the remarkable features of 
the $S$-matrices of real ATFTs is the fact that  
they display a weak-strong coupling duality, which means that the 
$S$-matrices in the simply-laced cases are invariant under the 
transformation 
\be 
\beta \lra \frac{4\pi}{\beta}\;. \label{duality} 
\ee 
It has been a long standing problem to understand whether a similar 
duality holds for the theories on a half-line. In a number of previous 
attempts to construct affine Toda reflection factors 
\cite{sasak93, kim95b, kim96, fring94}  it was assumed that the
theories on a 
half-line display the same weak-strong coupling duality, which means
the reflection factors were expected to be invariant under the
transformation (\ref{duality}). However, in a recent paper
\cite{corri97} it
was pointed out that the form of the breather reflection factors in
the sine--Gordon theory \cite{ghosh94b} implies that the reflection
factors in the sinh--Gordon case ($a_1^{(1)}$ ATFT) are not self-dual
in general. 
Similarly, in the $\ato$ affine Toda case we now find that 
none of the three reflection factors in (\ref{realref1}) is
self-dual.  However, it turns out that the two factors
$K_1^{(+)}(\th)$ and  $K^{(d)}_1(\th)$ are dual to each other, i.e.  
\be 
K^{(+)}_1(\th) \lra K^{(d)}_1(\th) \;, \hs{2cm} (\mbox{as}\;\;\; \beta
\lra  \frac{4\pi}{\beta})\;.  
\ee 
This supports the view expressed in \cite{corri97} that, rather than
being a symmetry of any particular boundary theory, duality should relate
theories with different boundary conditions to each other.   
 
In order to decide which boundary conditions could correspond to the 
reflection factors (\ref{realref1}) we first need to check their 
semiclassical limits. Expanding the reflection factors in terms of 
$\beta^2$ we obtain 
\bea 
K^{(+)}_1(\th) &=& 1 - \frac 14 \frac{\cos\(\frac{\th}{2i}\)} 
{\sin\(\frac{3\th}{2i}\)} \beta^2 + {\cal O}(\beta^4) \;, \nn \\ 
K^{(-)}_1(\th) &=& 1  - \frac 14 \frac{\sin\(\frac{\th}{2i}\)} 
{\cos\(\frac{3\th}{2i}\)} \beta^2 + {\cal O}(\beta^4)\;, \nn \\ 
K^{(d)}_1(\th) &=& - \bl -1 \br\, \bl -2 \br +  
\frac{\tan\(\frac{\th}{2i}\) \cos\(\frac{\th}{2i}+\frac{\pi}6\)}
{8\sin\(\frac{\th}{2i}+\frac{\pi}3\)
-4\cos\(\frac{3\th}{2i}+\frac{\pi}6\) +4\sin\(\frac{\th}{2i}\)}\,
\beta^2 + {\cal O}(\beta^4)\;. \nn \\       
\eea 
In \cite{kim95b} Kim used perturbative methods in order to construct
affine Toda reflection matrices for the Neumann boundary
condition. It turns out that his perturbative results\footnote 
{Note that the conjectured reflection factor in \cite{kim95b} is 
equal to $K_1^{(+)}(\th)$ only up to order $\beta^2$, since the exact
reflection factor in \cite{kim95b} was assumed to be self-dual.}  
for $\ato$ match the semiclassical limit of our $K^+_1(\th)$ up to
order $\beta^2$. We  
therefore conjecture that $K^+_1(\th)$ is the reflection factor for 
$\ato$ ATFT on a half-line with Neumann boundary condition. In terms
of the affine Toda boundary potential (\ref{todaB}) this
corresponds to the case where $b_0=b_1=b_2=0$.   
 
We can also compare our result with the paper \cite{corri94} in which 
Corrigan et al.\ conjectured the form of some of the reflection
factors in real coupling $\ato$ ATFT. In particular their `minimal'
conjecture for the reflection factor corresponding to the boundary
condition  with \mbox{$b_0=b_1=b_2= +2$} is equal to our
$K_1^{(d)}(\th)$. In a more recent paper \cite{perki98} Perkins and
Bowcock calculated the ${\cal O}(\beta^2)$ quantum corrections to the
classical reflection factor for this boundary condition. The result of
this computation was also found to be in agreement with the
semiclassical limit of our $K_1^{(d)}(\th)$.   
This now suggests that the theory with Neumann boundary condition is
dual 
to that with `positive' boundary condition $b_0 = b_1 = b_2 =
+2$, which matches the result obtained for the sine/sinh--Gordon
theories in \cite{corri97}.    
 
Unfortunately we do not have any simple explanation for the
remaining reflection factor $K_1^{(-)}(\th)$. The first problem is the
fact that the classical limit of  $K_1^{(-)}(\th)$ is equal to 1 just
like that of  $K_1^{(+)}(\th)$. However, classically only the reflection
factor corresponding to the Neumann boundary condition should be
unity. It therefore seems rather strange to have two different
reflection factors with this property. The other puzzle with
$K_1^{(-)}(\th)$ is the question of its dual. The dual of
$K_1^{(-)}(\th)$ 
is of course in itself a perfectly reasonable reflection factor which
satisfies all necessary constraints\footnote{In fact the dual of
$K_1^{(-)}(\th)$ is the `minimal' reflection factor conjectured in
\cite{corri94} to correspond to the boundary condition with
$b_0=b_1=b_2 = -2$.}, but we are not able to derive it from the
\uq{\ato} invariant reflection matrices.

Finally, let us briefly discuss the effect of the more general scalar
factors (\ref{Anew}).  These introduce an additional term
$k^{(\eta)}(\mu)$ of the form (\ref{keta}) into the lowest breather
reflection factors, which consequently introduces the following
additional term into the particle reflection factors:
\be 
k^{(\eta)}(\mu) \lra k_1^{(\eta)}(\th) \equiv \bl B\eta-\frac B2+\frac 
32\br\, \bl B\eta+\frac B2-\frac 12\br\, \bl -B\eta-\frac B2-\frac 
32\br\, \bl -B\eta+\frac B2+\frac 12\br\;.   
\ee 
We can check whether the inclusion of such a factor spoils the
boundary crossing and bootstrap conditions.  
It can be seen that this is not the case from the following short
computation:  
\be  
k_2^{(\eta)}(\th) \equiv k_1^{(\eta)}(\th+\frac{i\pi}3) 
k_1^{(\eta)}(\th-\frac{i\pi}3) =  \bl B\eta-\frac B2+\frac 
52\br\, \bl B\eta+\frac B2-\frac 32\br\, \bl -B\eta-\frac B2-\frac 
52\br\, \bl -B\eta+\frac B2+\frac 32\br\;, \label{k1k2}
\ee 
and thus 
\[ 
k_1^{(\eta)}(\th)k_2^{(\eta)}(i\pi-\th) = 1\;. 
\]   
In \cite{sasak93} Sasaki noticed that a reflection factor multiplied
by any function which satisfies the \mbox{$S$-matrix} bootstrap
equations 
is again a consistent reflection factor itself. In our case this
freedom is expressed in the possible inclusion of $k_1^{(\eta)}(\th)$,
since 
equation (\ref{k1k2}) is identical to the $S$-matrix bootstrap
equation of $\ato$ ATFT. Note that for $\eta = \frac 12 -
\frac 3{2B}$ the factor $k_1^{(\eta)}(\th)$ is just the $S$-matrix
$S_{11}(\th)$ itself. In general the inclusion of these factors
will violate condition (\ref{K1=K2}), since only if $\eta =
\frac{1+3m}B\;,$ (for any integer $m$), do we get $k_1^{(\eta)}(\th) =
k_2^{(\eta)}(\th)$. However, reflection factors which violate
(\ref{K1=K2}) are not necessarily ruled out since it was discovered in
\cite{bowco96} that it is possible to generalise the boundary
potential (\ref{todaB}) to include time derivatives of the
fields, in which case mass degenerate particles may no longer have the
same reflection factors.

Notice also that the inclusion of these factors does not provide us
with a solution to the problems regarding the factor
$K_1^{(-)}(\th)$. First we notice that the inclusion of a factor
$k_1^{(\eta)}$ does not change the classical limit of the reflection
factor, because 
\[
k_1^{(\eta)} \lra 1\;,\;\;\;\;\;\;\;\;\;\; (\mbox{for} \;\; \beta \to
0)\;.
\]
And secondly, it is not possible to find a value for $\eta$ such that
the reflection factor $K_1^{(-)}(\th)k_1^{(\eta)}(\th)$ would become
self-dual. We expect that the inclusion of any of these additional
factors can be ruled out, because they will introduce additional
physical strip poles which cannot be explained in terms of physically
allowed bound states.

\section{Discussion} 

The main results of this paper are contained in equations
(\ref{Kp}), (\ref{Km}) and (\ref{realref1}). We have found new
non-diagonal solutions to the BYBE and used them to construct
\uq{\ato} invariant reflection matrices. It was shown that these
reflection matrices satisfy the necessary consistency requirements of
boundary unitarity, boundary crossing and the boundary bootstrap.  
We then used the boundary bootstrap equations in order to derive 
reflection factors for the lowest breathers. 
Although the reflection matrices contained some free parameters, it
turned out that all these free parameters disappear in the breather
bootstrap and we end up with only a finite number of possible 
breather reflection matrices.  Since the breathers can be identified
with the real Toda quantum particles, this result is consistent with
the classical result that $a_2^{(1)}$ ATFT permits only a finite
number of integrable boundary conditions.
However, we only found three different reflection factors, whereas
there are nine integrable boundary conditions in the classical theory.
This fact is consistent with the possibility that not all boundary
conditions, which were found to be classically integrable, are also
quantum integrable\footnote{This would also confirm the work
\cite{fujii95} in which the authors show that some of the classically
integrable boundary conditions contain certain instabilities and
therefore do not lead to consistent quantum theories.}. 

One of the most interesting discoveries is the fact that none of the 
reflection factors turned out to be self-dual under the weak-strong  
coupling duality (\ref{duality}). Instead we confirmed the conjecture
that duality relates different boundary theories to each other. In
particular we found that the two reflection factors corresponding to
the Neumann and `positive' boundary conditions are dual to each other.  

Probably the most puzzling results concern the reflection factor 
$K_1^{(-)}(\th)$ in (\ref{realref1}). This factor has a classical limit
of one, just like the factor $K_1^{(+)}(\th)$, which corresponds to the
Neumann boundary condition. However, we would expect that only the
reflection factor corresponding to the Neumann boundary 
condition should be one in the classical limit. 
Is the second factor $K_1^{(-)}(\th)$ not related to real coupling ATFT 
but to some other as yet unknown theory? Or could it be possible that
two different quantum theories correspond to the same classical
boundary condition? 

The other problem concerning the factor $K_1^{(-)}(\th)$ is the fact
that we were not able to find a solution to the BYBE which would lead to
the dual of $K_1^{(-)}(\th)$. 
However, we have no reason to assume that
the weak-strong coupling duality must be present in the 
solutions to the BYBE. Even in the bulk theory it was
found that weak-strong coupling duality does not exist in the
trigonometric $S$-matrices, but appears somewhat mysteriously 
only on  the level of the lowest breathers \cite{gande95, gande95b}.  

Apart from trying to understand these issues concerning the factor 
$K_1^{(-)}(\th)$ there are a number of other open questions. In
particular, it would be interesting to study the pole structure of our
reflection matrices and examine the appearance of possible boundary
bound states. Furthermore, in the solutions to the BYBE we have not
allowed the possibility of different boundary states. If the boundary
itself carried some quantum number then the $K$-matrices would depend
on two additional indices, i.e.  
\[ 
A_i(\mu)\cB_a = K_{i,a}^{j,b}(\mu)\,A_j(-\mu)\cB_b\;. 
\]  
In this case the BYBE would become a great deal more complicated and
at this stage we are not able to provide a general solution. 
We do not know what effect the inclusion of these boundary states
would have on the final results. However, it is interesting to note
that in the sine--Gordon theory the solution of the simplest BYBE
(i.e.\ that without boundary labels) seems to give all possible
breather reflection factors.  

Another open problem is the extension of this work to other
algebras. We expect to perform a similar analysis for the case of
$\att$ ATFT in a forthcoming paper. The case of $\att$ is interesting, 
because unlike in the $\ato$ theory we expect to find at least
one free boundary parameter in the lowest breather reflection
factors \cite{bowco95}. 
Apart from the theories based on $\ato$ and $\att$, it is
currently not possible to construct reflection matrices for any other
algebra. The reason for this is that we do not 
know the exact matrix structure of any of the higher trigonometric 
$R$-matrices, which are usually given only in terms of their spectral  
decompositions. Rather than solving the BYBE step by step, it would be
highly desirable to find a more general construction in order to find
solutions. 
We hope that the explicit solutions in this paper will eventually 
lead to a better understanding of the general structure of solutions
to the BYBE and thus provide a more general method for 
constructing reflection matrices associated with
higher rank algebras.

\vs{2.5cm} 
 
\noi {\bf \ul{Acknowledgements:}}\\ 
I would like to thank Peter Bowcock, Ed Corrigan, Gustav Delius,
Patrick Dorey and Davide Fioravanti for discussions during the course
of this work. I am supported by an EPSRC research grant no.\ $GR/K\, 
79437\,$.

\newpage

\appendix 
\section{Solving the Boundary Yang--Baxter--Equation}   

In this appendix we present some of the steps\footnote{All the
computations were done using MapleV. We present all the results here
in terms of $\mu$ and $\la$.  
However, for technical reasons most of the computation have been 
performed using the variables $x=e^{i\pi\mu/3}$ and $q=e^{i\pi\la/4}$, 
since in this case all trigonometric equations can be written in terms 
of simpler rational equations.} 
in the solution of the \uq{\ato} invariant BYBE.
We only show the solution for $K^{\pm}(\mu)$ (the solution for the
conjugate reflection matrices $\ol K^{\,\pm}(\mu)$ is completely
analogous).     
 
The BYBE was given in explicit matrix form in equation (\ref{BYBE})
and illustrated in {\em figure 2}, in which two incoming
solitons $A_i(\mu)$, $A_j(\mup)$ reflect into two outgoing solitons
$\ol  A_p(-\mu)$, $\ol A_r(-\mup)$.    
Hence, relation (\ref{BYBE}) leads to 81 independent equations,
which we label by $(i,j,p,r)$, for the nine unknown functions
$K_k^l(\mu)$, which are the matrix elements of the reflection matrices
$K^{\pm}(\mu)$.  
For the sake of simplicity we can remove the overall $S$-matrix scalar
factors from the BYBE, by dividing both sides of   
(\ref{BYBE}) by $\frac{F(\mu-\mup)}{\sin(\pi(\la-\mu+\mup))}   
\frac{F(\frac32 \la-\mu-\mup)}{\sin(\pi(-\frac12   
\la+\mu+\mup))}$, which is the same as taking 
\bea   
S_{i,i}^{i,i}(\mu) &=& \sin(\pi(\la-\mu))\;,    
\nn \\   
S_{i,j}^{j,i}(\mu) &=&  \sin(\pi\mu)\;,    
\nn \\   
S_{i,j}^{i,j}(\mu) &=& \sin(\pi\la)e^{\nu(i,j)i\pi\frac{\mu}3}\;,    
\;\;\;\;\;\;\; (\mbox{for}\;\;\; i,j = 1, \dots, 3) \;,   
\eea   
instead of (\ref{Smcomp}).   
 
First let us consider equation (1,1,2,2) which only depends on the   
diagonal elements $K_1^1$ and $K_2^2$. Assuming neither of the 
diagonal elements is identically zero, this equation can be written in 
the very simple form    
\[   
\frac{K_1^1(\mu)}{K_2^2(\mu)} = \frac{K_1^1(\mup)}{K_2^2(\mup)}\;,     
\]   
which implies that $K_1^1$ and $K_2^2$ have the same rapidity   
dependence. We can also find similar equations involving $K_3^3$ and
we thus define:    
\bea   
K_2^2(\mu) &=& \cD_2(\la)K_1^1(\mu)\;, \nn \\   
K_3^3(\mu) &=& \cD_3(\la)K_1^1(\mu)\;,    
\eea   
in which $\cD_2(\la)$ and $\cD_3(\la)$ are two arbitrary functions   
depending on the coupling constant $\la$ (but not on the rapidity   
$\mu$).   
    
Next we consider equation (1,2,1,2), which only depends on the two   
functions $K_1^2$ and $K_2^1$, and which can be simplified to   
\bea   
\lefteqn{\sin\(\pi(\mu-\mup)\)   
\left[K_1^2(\mu)K_1^2(\mup)e^{i\pi(\frac{\la}2-\frac{\mu+\mup}3)}    
-K_2^1(\mu)K_2^1(\mup)e^{-i\pi(\frac{\la}2-\frac{\mu+\mup}3)}\right]   
=} \hs{3cm} \nn \\   
& & = \sin\(\pi(\mu+\mup-\frac{\la}2)\)   
\left[K_1^2(\mu)K_2^1(\mup)e^{i\pi\frac{\mup-\mu}3}    
-K_2^1(\mu)K_1^2(\mup)e^{i\pi\frac{\mu-\mup}3}\right]\;. \nn   
\eea   
Apart from the trivial solution in which both $K_1^2$ and $K_2^1$ are 
zero, there are two other possible solutions to this equation for 
general $\mu$ and $\mup$, namely either 
\bea 
K_1^2(\mu) &=& e^{-\frac 23 i\pi\mu}\cA_1(\mu)\;, \nn \\ 
K_2^1(\mu) &=& e^{\frac 23 i\pi\mu}\cA_1(\mu)\;,  
\eea  
or 
\bea   
K_1^2(\mu) &=& e^{i\pi(\frac{\mu}3-\frac{\la}4)}\cA_1(\mu)\;, \nn   
\\   
K_2^1(\mu) &=& \ep_1 e^{-i\pi(\frac{\mu}3-\frac{\la}4)}\cA_1(\mu)\;,
\label{K12}    
\eea   
in which $\ep_1 = \pm 1$ and $\cA_1(\mu)$ is an arbitrary function of   
$\mu$ and $\la$. However, after a careful analysis of the other
equations we find that the first of these solutions 
inevitably leads to contradictions and we are therefore left with
solution (\ref{K12}).     
The corresponding equations for the other non-diagonal elements are
(1,3,1,3) and (2,3,2,3), and we can make the following ansatz:
\bea   
K_2^3(\mu) &=& e^{i\pi(\frac{\mu}3-\frac{\la}4)}\cA_2(\mu)\;, \nn   
\\   
K_3^2(\mu) &=& \ep_2 e^{-i\pi(\frac{\mu}3-\frac{\la}4)}\cA_2(\mu)\;,  
\eea   
and   
\bea   
K_3^1(\mu) &=& e^{i\pi(\frac{\mu}3-\frac{\la}4)}\cA_3(\mu)\;, \nn   
\\   
K_1^3(\mu) &=& \ep_3 e^{-i\pi(\frac{\mu}3-\frac{\la}4)}\cA_3(\mu)\;.  
\eea   
So far we have reduced the nine unknown functions $K_i^j(\mu)$ to
four $\mu$ dependent functions $K_1^1(\mu)$, $\cA_1(\mu)$, $\cA_2(\mu)$,  
$\cA_3(\mu)$, two $\la$ dependent functions $\cD_2(\la)$,
$\cD_3(\la)$, and the choices of signs $\ep_1,\ep_2,\ep_3$.   
  
Putting the above ansatz back into the BYBE we find that, for example,
equation (1,1,1,2) now only depends on $\cA_1$ and $K_1^1$ and can be 
simplified to  
\[  
K_1^1(\mu)\cA_1(\mup) \sin(\pi(2\mup-\frac{\la}2)) =   
K_1^1(\mup)\cA(\mu) \left[ \sin(\pi(\mu+\mup-\frac{\la}2)) + \ep_1  
\sin(\pi(\mu-\mup)) \right]\;.  
\]   
We can separate the $\mu$ and $\mup$ dependence in this equation and
write  
\bea  
\frac{\cA(\mu)}{K_1^1(\mu)}\sin(\pi(\mu-\frac{\la}4)) &=&   
\frac{\cA(\mup)}{K_1^1(\mup)}\sin(\pi(\mup-\frac{\la}4))\;,  
\hs{1cm} (\mbox{if}\;\; \ep_1 = +1)\;, \nn \\ \nn \\  
\frac{\cA(\mu)}{K_1^1(\mu)}\cos(\pi(\mu-\frac{\la}4)) &=&   
\frac{\cA(\mup)}{K_1^1(\mup)}\cos(\pi(\mup-\frac{\la}4))\;,  
\hs{1cm} (\mbox{if}\;\; \ep_1 = -1)\;. \nn  
\eea  
This can be solved by introducing yet another $\la$ dependent
parameter $\al(\la)$:
\be  
K_1^1(\mu) = \al(\la) \sin\(\pi(\mu-\frac{\la}4+\frac14  
(1-\ep_1))\)\, \cA_1(\mu)\;.   
\ee  
Considering equation (1,2,1,3) we then obtain  
\bea  
\lefteqn{\cA_1(\mu)\cA_3(\mup) \ep_3 \left[    
\ep_1\sin(\pi(\mu+\mup-\frac{\la}2)) + \sin(\pi(\mu-\mup)) \right]    
=} \hs{3cm} \nn \\  
&& = \cA_1(\mup)\cA_3(\mu) \ep_1 \left[   
\ep_3\sin(\pi(\mu+\mup-\frac{\la}2)) + \sin(\pi(\mu-\mup)) 
\right]\;. \label{sol3}     
\eea  
However from  equations (1,1,2,3) we find that we must have 
\[ 
\ep_1 = \ep_2 = \ep_3 \equiv \ep\;,
\]
which simplifies (\ref{sol3}) to  
\[  
\frac{\cA_1(\mu)}{\cA_3(\mu)} = \frac{\cA_1(\mup)}{\cA_3(\mup)}\;.  
\] 
We can find a similar relation (from equation (2,3,2,1)) for the
function $\cA_2(\mu)$ and we thus make the ansatz  
\bea 
\cA_2(\mu) = g(\la)\cA_1(\mu)\;, \nn \\ 
\cA_3(\mu) = h(\la)\cA_1(\mu)\;,  
\eea 
in which $h(\la)$ and $g(\la)$ are again arbitrary functions of $\la$.

Now we are left with one $\mu$ dependent function $\cA_1(\mu)$,    
five $\la$ dependent parameters $\cD_2(\la)$, $\cD_3(\la)$,
$\al(\la)$, $h(\la)$ and $g(\la)$, and one choice of sign $\ep$. If we
put all these back into the BYBE we find that only three independent
non-trivial equations remain (namely (1,1,2,3),  
(1,2,2,3) and (1,2,3,3)\,), which for the case of $\ep =+1$ can be  
written as 
\bea 
h(\la) &=& \sin(\pi\frac{\la}2)\al(\la)g(\la)\;, \nn \\ 
g(\la) &=& \sin(\pi\frac{\la}2)\al(\la)h(\la)\cD_2(\la)\;, \nn \\  
h(\la)g(\la) &=& \sin(\pi\frac{\la}2)\al(\la)\cD_3(\la)\;, \nn
\eea 
and therefore 
\bea 
\al(\la) &=& \frac1{\sin(\pi\frac{\la}2)} \frac{h(\la)}{g(\la)}\;, \nn 
\\ 
\cD_2(\la) &=& \(\frac{g(\la)}{h(\la)}\)^2\;, \nn \\ 
\cD_3(\la) &=& \(g(\la)\)^2\;. \nn 
\eea 
The only difference in the case of $\ep =-1$ is  
\[ 
\al(\la) =  \frac i{\sin(\pi\frac{\la}2)} \frac{h(\la)}{g(\la)}\;. 
\] 
Now, putting everything together we obtain the two non-diagonal
solutions (\ref{BYBEsol1}) to the BYBE as given in section 3. The
reflection matrices for the reflection of antisolitons into solitons
can be obtained in a completely analogous way by using equation
(\ref{BYBE2}). 
In all these solutions we have assumed that none of the matrix elements
are identically zero. In a careful case by case study one can see that
the only possibility of having a solution with any matrix elements
being identically zero is the diagonal solution (\ref{diagsol}). We
therefore expect that the three types of
solutions given in section 3 are indeed the most general solutions to
the BYBE (\ref{BYBE}).

\vs{0.3cm}

Finally we briefly want to mention possible solutions to a slightly
different BYBE which, nevertheless, is also associated to the quantum
algebra \uq{\ato}. Unlike previously thought, $K$-matrices of the form
(\ref{Kmatrices}) are not the only possible solutions to the BYBE. 
There are in fact also reflection matrices which map each multiplet
into itself \footnote{I am grateful to Rafael
Nepomechie for bringing this to my attention.}:
\be
\tK (\mu): V_1 \to V_1\;. \label{KV1V1} 
\ee 
Reflection matrices of this type have to satisfy the following BYBE:
\be  
\tK_j^k(\mup)\, S_{i,k}^{l,m}(\mu+\mup)\, \tK_m^n(\mu)\, 
S_{l,n}^{p,r}(\mu-\mup) =  S_{i,j}^{k,l}(\mu-\mup)\, \tK_l^m(\mu)\, 
S_{k,m}^{p,n}(\mu+\mup)\, \tK_n^r(\mup)\;. \label{BYBEapp}  
\ee  
The most general diagonal solutions to this equation (and its
generalisation to $a_n^{(1)}$) have been found some time ago in
\cite{deveg93}.   
Up to an arbitrary overall scalar factor these are
\be
\tK_j^k(\mu) = 0\;,\;\;\;\;\;\;\;\;\; (\mbox{for} \;\;\; j \neq k)\;,
\ee 
and either
\be
\tK_j^j(\mu) = e^{i\pi\frac 43 \mu (j-1)}\;, \;\;\;\;\; (j =1,2,3)\;,
\label{tildeK1} 
\ee 
or
\bea
\tK_j^j(\mu) &=& e^{i\pi\frac 43 \mu (j-1)} \tilde{a}(\mu)\;, \nn \\
\tK_k^k(\mu) &=& e^{i\pi\frac 43 \mu (k-1)} \tilde{b}(\mu)\;,
\;\;\;\;\;\; (1\leq j < k \leq 3) \label{tildeK2}
\eea
in which
\be
\tilde{a}(\mu) = e^{i\pi\mu} \sin(\pi(\mu+\xi))\;, \;\;\;\;\;\;\;\;\;
\tilde{b}(\mu) = e^{-i\pi\mu} \sin(\pi(\xi-\mu))\;,
\ee
and $\xi$ is a free boundary parameter. 
Note that the difference between (\ref{tildeK1}, \ref{tildeK2}) and the
expressions in \cite{deveg93} are due to the fact that in
\cite{deveg93} the $R$-matrix in the homogeneous 
gradation was used whereas here we are working with the $R$-matrix in
the principal gradation. This means that the solutions to the
corresponding BYBE are related to each other by a simple `gauge'
transformation, which leads to the additional factor $e^{i\pi\frac 43
\mu(j-1)}$ in the above expressions.

These $K$-matrices have been used in the study of open quantum spin
chains (see for instance \cite{doiko98}). However, as mentioned
earlier, because of the fact that affine Toda solitons change
multiplets after reflection, we do not expect the reflection matrices
(\ref{tildeK1}, \ref{tildeK2}) to be relevant in the study of affine
Toda field theories.

\section{Evaluating the boundary bootstrap relations} 
\subsection{The soliton bootstrap equations} 

In this appendix we provide the details of the soliton bootstrap
relations which were illustrated in {\em figure 3} in section 5.1.
We make explicit use of the FZ algebra which was defined in
(\ref{FZAi}), (\ref{FZrefl}) and (\ref{solfuse}). In order to avoid
any confusion we write out explicitely all summations over indices in
this appendix.  

The soliton bootstrap involves the reflection of two solitons from the
boundary, and we therefore write 
\be 
A_j(\mu_1)\, A_k(\mu_2) \cB = \sum_{m,n,p,r=1}^3 K_k^p(\mu_2)\,
S_{j,\ol p}^{\ol m,r}(\mu_1+\mu_2)\, K_r^n(\mu_1)\, \ol A_m(-\mu_2)\,
\ol A_n(-\mu_1)\, \cB\;,   
\ee 
which describes the reflection of two incoming solitons $A_j, A_k$
into two outgoing solitons $\ol A_m,\ol A_n$. For later convenience
we define    
\be 
H_{j,k}^{m,n} \equiv  \sum_{p,r=1}^3 K_k^p(\mu+\frac{\la}2)\, 
S_{j,\ol p}^{\ol m,r}(2\mu)\, K_r^n(\mu-\frac{\la}2)\;.  
\ee  
Using the explicit expression for the $S$-matrix and $K$-matrices we
can check the following identity, which is true for all $j \neq k$: 
\be 
H_{j,k}^{m,n} + \ga_{j,k} H_{k,j}^{m,n} = \left\{ 
\begin{array}{ll}  
\ga_{m,n} \left[ H_{j,k}^{n,m} + \ga_{j,k} 
H_{k,j}^{n,m}\right]\;,\;\;\;\;\; &   
(\mbox{if} \;\;\;\; m\neq n) \\ \\
0\;, &  (\mbox{if} \;\;\;\; m = n)\;. \end{array} \right. 
\ee 
These identities will turn out to be necessary conditions for the
soliton  bootstrap relations to be satisfied.       
Now let $(i,j,k)$ be an even permutation of $(1,2,3)$, and define $\mu
= \frac{\mu_1+\mu_2}2\;, \;\; \mu_1 = \mu-\frac{\la}2\;,$ and $\mu_2 =
\mu+\frac{\la}2$.  Then using (\ref{solfuse}) and the above identities
we can write  
\bea 
\ol A_i(\mu)\, \cB &=& \left[ A_j(\mu_1)\, A_k(\mu_2) + \ga_{j,k} 
A_k(\mu_1)\, A_j(\mu_2)\right]\, \cB  \nn \\  
&=& \sum_{m,n = 1}^3 \( H_{j,k}^{m,n} + \ga_{j,k} 
H_{k,j}^{m,n} \) \ol A_m(-\mu_2) \ol A_n(-\mu_1)\, \cB \nn \\ 
&=& \sum_{(m,n) = (2,1),(1,3),(3,2)}  \( H_{j,k}^{m,n} + \ga_{j,k}  
H_{k,j}^{m,n} \) \ol A_m(-\mu_2) \ol A_n(-\mu_1)\, \cB \nn \\ 
&& + \sum_{(m,n) = (2,1),(1,3),(3,2)}  \( H_{j,k}^{n,m} + \ga_{j,k}  
H_{k,j}^{n,m} \) \ol A_n(-\mu_2) \ol A_m(-\mu_1)\, \cB \nn \\ 
&& + \sum_{m=1}^3 \( H_{j,k}^{m,m} + \ga_{j,k} 
H_{k,j}^{m,m} \) \ol A_m(-\mu_2) \ol A_m(-\mu_1)\, \cB \nn \\  
&=& \sum_{(m,n) = (2,1),(1,3),(3,2)}  \( H_{j,k}^{m,n} + \ga_{j,k} 
H_{k,j}^{m,n} \) \( \ol A_m(-\mu_2) \ol A_n(-\mu_1) + \ga_{n,m} \ol 
A_n(-\mu_2) \ol A_m(-\mu_1)\)\, \cB \nn \\ 
&=& \sum_{(l,n,m) = (1,2,3),(3,1,2),(2,3,1)}  \( H_{j,k}^{m,n} + 
\ga_{j,k} H_{k,j}^{m,n} \)  A_l(-\mu)\, \cB\;. \nn
\eea 
Hence, we find that the soliton bootstrap relations imply
\be 
\ol K_i^{\,l}(\mu) = H_{j,k}^{m,n} + \ga_{j,k}\,H_{k,j}^{m,n}\;, 
\label{solboot1}   
\ee 
in which 
$(i,j,k)$ and $(l,n,m)$ are even permutations of $(1,2,3)$. 
 For the sake of clarity let us consider one specific example here,
namely  $i=1, \;\; l = 2\;$ and $\; \ep = +\;$:  
\bea 
H_{2,3}^{1,3} + \ga_{2,3}\,H_{3,2}^{1,3} &=& \sum_{p,r=1}^3 \,\left[ 
K_3^p(\mu+\frac{\la}2)\,    
S_{2,\ol p}^{\ol 1,r}(2\mu)\, K_r^2(\mu-\frac{\la}2) + e^{i\pi\frac23} 
K_2^p(\mu+\frac{\la}2)\, S_{3,\ol p}^{\ol 1,r}(2\mu)\, 
K_r^2(\mu-\frac{\la}2)\right] = \nn \\   
&=& -g(\la)h(\la)\, e^{-i\pi(\frac{\mu}3 - 
\frac{\la}4)}\,   \frac {\sin\(\pi(\mu+\frac{\la}4)\)  
\sin\(\pi(\mu-\frac 54 {\la})\) \cos\(\pi(\mu-\frac 34 {\la})\)} 
{\sin\(\pi\frac{\la}2\) \sin\(\pi(\mu-\frac{\la}4)\) 
\cos\(\pi(\mu-\frac{\la}4)\)}\nn \\ && \times \cA^{+}(\mu+\frac{\la}2)
F(\frac32 {\la}  - 2\mu) \cA^{+}(\mu-\frac{\la}2) \;. 
\label{solboot2}     
\eea 
First we need to compute the overall scalar factor on the right hand 
side. 
After some lengthy calculations involving the usual Gamma function 
manipulations we obtain the following identities:   
\[ 
a_1(\mu+\frac{\la}2)\,a_1(\mu-\frac{\la}2)\, F(\frac 32 
\la-2\mu) =  \frac{\sin\(\pi(2\mu-\frac{\la}2)\)}{\sin\(\pi(2\mu-\frac32 
\la)\)}\, \bl \frac{\la}4 \br \bl -\frac34 \la \br \bl \frac54 \la 
\br\, a_1(\mu)\;,
\]
and  
\bea 
a^+_0(\mu-\frac{\la}2)\,a^+_0(\mu-\frac{\la}2) &=& 
\frac{\sin\(\pi(\mu-\frac34 \la)\)}{\sin\(\pi(\mu-\frac54 \la)\) 
\sin\(\pi(\mu+\frac{\la}4)\)}\, \bl \frac{\la}4 \br \bl -\frac34 \la 
\br \bl \frac54 \la \br\, a^+_0(\mu)\;, \nn \\ \nn \\
a^-_0(\mu-\frac{\la}2)\,a^-_0(\mu-\frac{\la}2) &=& 
\frac{\cos(\pi(\mu-\frac34 \la))}{\cos(\pi(\mu-\frac54 
\la))\cos(\pi(\mu+\frac{\la}4))}\, \bl -\frac{\la}4 \br \bl \frac34 
\la \br \bl -\frac54 \la \br\, a^-_0(\mu)\;. \nn 
\eea  
And these lead to 
\bea 
\cA^+(\mu+\frac{\la}2) F(\frac 32 \la - 2\mu) 
\cA^+(\mu-\frac{\la}2) &=& \frac{\sin\(\pi\frac{\la}2\) 
\sin\(\pi(\mu-\frac{\la}4)\) \cos\(\pi(\mu-\frac{\la}4)\)} 
{\sin\(\pi(\mu+\frac{\la}4)\) \sin\(\pi(\mu-\frac 54 {\la})\) 
\cos\(\pi(\mu-\frac 34 {\la})\)}\nn \\ &&  \times \bl \frac{\la}4
\br^2 \bl -\frac34 \la \br^2 \bl \frac54 \la \br^2\,
\ol \cA^{\,+}(\mu)\;, \nn \\ \nn \\   
\cA^-(\mu+\frac{\la}2) F(\frac 32 \la - 2\mu) \cA^-(\mu-\frac{\la}2)
&=& - \frac{\sin\(\pi\frac{\la}2\) \sin\(\pi(\mu-\frac{\la}4)\) 
\cos\(\pi(\mu-\frac{\la}4)\)} {\cos\(\pi(\mu+\frac{\la}4)\) 
\cos\(\pi(\mu-\frac 54 {\la})\) \sin\(\pi(\mu-\frac 34 {\la})\) 
}\,\ol \cA^{\,-}(\mu)\;, \nn \\ \nn \\
\cA^d(\mu+\frac{\la}2) F(\frac 32 \la - 2\mu) \cA^d(\mu-\frac{\la}2)
&=&  \frac{\sin\(\pi(2\mu-\frac{\la}2)\)}{\sin\(\pi(2\mu-\frac32 
\la)\)}\, \ol \cA^{\,d}(\mu)\;. \nn
\eea   

From this we can see that only in the case of $\cA^+$ we have to
include an additional CDD-type factor $\si(\mu)$, which satisfies   
equations (\ref{sigma1}) and  
\be 
\si(\mu+\frac{\la}2)\si(\mu-\frac{\la}2) =  \bl -\frac{\la}4 \br^2 \bl 
\frac34 \la \br^2 \bl -\frac54 \la \br^2\, \si(\mu)\;.  
\ee 
It is easy to check that a solution to these equations if provided by
$\si^+(\mu)$ which was given in (\ref{sigmap}).

Now we can go back to the above example and we find that relation 
(\ref{solboot2}) simplifies to  
\[ 
H_{2,3}^{1,3} + \ga_{2,3}\,H_{3,2}^{1,3} = -g(\la)h(\la)\, \ol 
K_1^{\,2}(\mu)\;. 
\] 
Hence in order for the bootstrap to be satisfied we require that
\[
g(\la)h(\la) = -1\;.
\]
It is easily verified that this restriction is sufficient to ensure
that (\ref{solboot1}) is true for all $\;i,l = 1,2,3$ (with $i\neq l$).  
We can also check that similar relations hold for the cases of $\;\ep =
-\;$ and $\;\ep = d\,$, and we find that the soliton bootstrap
equations are 
all satisfied if the free parameters in the $K$-matrices are
restricted by (\ref{solrestr}). Therefore, we obtain the full
reflection matrices (\ref{Kp}) and (\ref{Km}), which were given in
section 5.1.

\subsection{The breather bootstrap equations}  

Using the exchange relations (\ref{FZAi}) and (\ref{FZrefl}) 
we can write the scattering process on the right hand side of  
{\em figure 4} in the following form:  
\be 
A_m(\mu_1) \ol A_m(\mu_2)\cB = \sum_{j,k,l,n = 1}^3 \ol 
K_m^{\,j}(\mu_2) S_{m,j}^{k,l}(\mu_1+\mu_2) K_l^n(\mu_1)\,
A_k(-\mu_2)\ol  A_n(-\mu_1) \cB\;.  
\ee 
Let us therefore introduce the abbreviations 
\[ 
G_m^{k,n}(\mu) \equiv \sum_{j,l=1}^3 \ol K_m^{\,j}(\mu+\frac 34
\la-\frac12)\, S_{m,j}^{k,l}(2\mu)\, K_l^n(\mu-\frac 34
\la+\frac12)\;,  
\] 
and
\[ 
J^{k,n}(\mu) \equiv \sum_{m=1}^3 \al_m G_m^{k,n}(\mu)\;. 
\] 
Then we can check that the following identities hold: 
\be 
J^{k,n}(\mu) = 0\;,\;\hs{2cm} (\mbox{if} \;\; k \neq n)\;,
\ee
and
\be 
\hs{-1cm} J^{1,1}(\mu) = \al_1\, J^{2,2}(\mu) = \al_3\, J^{3,3}(\mu)\;. 
\ee 
Using these identities and the FZ generators for the lowest
breathers (\ref{FZB}), it is now straightforward to express the
breather reflection amplitude in terms of the FZ algebra:        
\bea 
B_1(\mu)\,\cB &=& \sum_{m=1}^3 \al_m\, A_m(\mu-\frac 34 \la+\frac12)\, 
\ol A_m(\mu+\frac 34 \la-\frac12)\,\cB  \nn \\ 
&=& \sum_{m=1}^3 \al_m \sum_{k,n = 1}^3 G_m^{k,n}(\mu)\, 
A_k(-\mu-\frac 34 \la+\frac12)\, \ol A_n(-\mu+ \frac 34 \la 
-\frac12)\, \cB  \nn \\  
&=& \sum_{k,n = 1}^3 J^{k,n}(\mu)\, A_k(-\mu-\frac 34 \la+\frac12)\, 
\ol A_n(-\mu+ \frac 34 \la -\frac12)\, \cB  \nn \\ 
&=&  \sum_{k = 1}^3 J^{k,k}(\mu)\,  A_k(-\mu-\frac  
34 \la+\frac12)\, \ol A_k(-\mu+ \frac 34 \la -\frac12)\, \cB  \nn \\ 
&=&  \al_1^{-1} J^{1,1}(\mu) \sum_{k=1}^3 \al_k\,  A_k(-\mu-\frac  
34 \la+\frac12)\, \ol A_k(-\mu+ \frac 34 \la -\frac12)\, \cB  \nn \\  
&=& \al_1^{-1} J^{1,1}(\mu)\,\, B_1(-\mu)\, \cB\;, \nn \\ \nn
\eea 
We therefore define the lowest breather reflection factor as
\be 
K_{B}(\mu) \equiv \al_1^{-1} J^{1,1}(\mu)\;, 
\ee 
which can now be computed explicitely, and we obtain 
\[ 
\al_1^{-1}\, J^{1,1}(\mu) = 
\cA^{\ep}(\mu-\frac34 \la+\frac12)\, F(2\mu)\, \ol
\cA^{\,\ep}(\mu+\frac34 \la- \frac12)\, E^{\ep}(\mu)  \;, 
\] 
in which  
\be 
E^{\ep}(\mu) = \left\{ \begin{array}{cl} 
- \frac{\cos(\pi\mu) \cos\(\pi(\mu+\frac{\la}2)\) \cos\(\pi(\mu-\frac
32\la)\)} {\sin^2\(\pi\frac{\la}2\)
\cos\(\pi(\mu-\frac{\la}2)\)}\;,\;\;\;\; & (\mbox{if}\;\;\; 
\ep = +)\;, \\ \\
- \frac{\sin(\pi\mu) \sin\(\pi(\mu+\frac{\la}2)\)
\sin\(\pi(\mu-\frac 32\la)\)} {\sin^2\(\pi\frac{\la}2\)
\sin\(\pi(\mu-\frac{\la}2)\)} 
\;, & (\mbox{if}\;\;\; \ep = -)\;, \\ \\
1 \;, & (\mbox{if}\;\;\; \ep = d)\;. 
\end{array} \right. \label{E}
\ee
And after some straightforward computations involving the 
scalar factors we find    
\bea 
\cA^{+}(\mu-\frac34 \la+\frac12)\, 
F(2\mu)\, \ol \cA^{\,+}(\mu+\frac34 \la- \frac12) &=&  - \frac 
{\cos\(\pi(\mu-\frac{\la}2)\) \sin^2\(\pi\frac{\la}2\)} {\cos(\pi\mu)  
\cos\(\pi(\mu+\frac{\la}2)\) \cos\(\pi(\mu-\frac 32\la)\)} 
\nn \\ && \times 
\bl \frac 12\br\, \bl -\la \br \, \bl -2\la+\frac 12\br\,  
\bl \frac{\la}2 -\frac 12\br\, \bl -\frac{\la}2 -\frac 12\br\;, \nn \\
\nn \\ 
\cA^{-}(\mu-\frac34 \la+\frac12)\, 
F(2\mu)\, \ol \cA^{\,-}(\mu+\frac34 \la- \frac12) &=&  - \frac 
{\sin\(\pi(\mu-\frac{\la}2)\) \sin^2\(\pi\frac{\la}2\)} {\sin(\pi\mu)  
\sin\(\pi(\mu+\frac{\la}2)\) \sin\(\pi(\mu-\frac 32\la)\)} \nn \\ 
&& \times  \bl  
\frac{\la}2 \br\, \bl -\frac{\la}2-\frac 12 \br \, \bl -\frac 32 \la+ 
\frac 12\br\;, \nn \\ \nn \\  
\cA^{d}(\mu-\frac34 \la+\frac12)\, 
F(2\mu)\, \ol \cA^{\,d}(\mu+\frac34 \la- \frac12) &=&   \bl
-\frac{\la}2 - \frac 12 \br\, \bl -\la \br\,  \bl -\frac 32 \la +
\frac 12\br\;,  \nn
\eea 
from which we can see that all the terms in (\ref{E}) get cancelled. 
Finally, in the case of $K^+(\mu)$ we also have to take the
contribution from the additional CDD factor (\ref{sigmap}) into
account, i.e.\
\[
\si^+(\mu-\frac34 \la+\frac12)\, \si^+(\mu+\frac34 \la- \frac12) = 
 \bl \la-\frac 12\br\, \bl 2\la-\frac 12 \br\,  \bl -\frac{\la}2 +
 \frac 12\br\, \bl \frac{\la}2 + \frac 12\br\;.  
\]
And, therefore, we obtain the final result, the reflection factors for 
the reflection of a breather $B_1(\mu)$ from the boundary 
\be 
B_1(\mu) \cB = K^{(\ep)}_B(\mu) B_1(-\mu) \cB\;, 
\ee 
in which 
\bea 
K^{(+)}_B(\mu) &=& \bl \frac 12\br\, \bl -\la \br\,  
\bl \la - \frac 12\br\;, \nn \\ \nn \\
K^{(-)}_B(\mu) &=& - \bl \frac{\la}2\br\, \bl -\frac{\la}2-\frac 12 \br\,  
\bl -\frac 32 \la + \frac 12\br\;, \nn \\ \nn \\
K^{(d)}_B(\mu) &=&   \bl -\frac{\la}2 - \frac 12 \br\, \bl -\la \br\,  
\bl -\frac 32 \la + \frac 12\br\;. \nn
\eea

\vs{1.5cm}


\parskip 1pt       
{\footnotesize        
\begin{thebibliography}{99}        
%
\bibitem{corri94} E.\ Corrigan, P.E.\ Dorey, R.H.\ Rietdijk and R.\ 
Sasaki, {\em Affine Toda field theory on a half-line}, 
Phys.Lett.\ {\bf B333} (1994), 83; {\tt hep-th/9404108};\\  
E.\ Corrigan, P.E.\ Dorey and R.H.\ Rietdijk, 
{\em Aspects of affine Toda field theory on a half-line}, 
Prog.Theor.Phys.Suppl.\ {\bf 118} (1995), 143; {\tt hep-th/9407148}  
%
\bibitem{bowco95} P.\ Bowcock, E.\ Corrigan, P.E.\ Dorey and R.H.\
Rietdijk, {\em Classically integrable boundary conditions for affine
Toda field theories}, Nucl.Phys.\ {\bf B445} (1995), 469; {\tt
hep-th/9501098}  
%
\bibitem{corri96} E.\ Corrigan, {\em Integrable Field Theory with     
Boundary Conditions}, preprint DTP--96/49, {\tt hep-th/9612138}     
%
\bibitem{chere84} I.V.\ Cherednik, {\em Factorizing particles on a
half-line and root systems}, Theor.Math.Phys.\ {\bf 61} (1984), 977
%
\bibitem{sklya88} E.K.\ Sklyanin, {\em Boundary conditions for
integrable equations}, Jour.Phys.\ {\bf A21} (1988), 2375 
%
\bibitem{ghosh94} S.\ Ghoshal and A.\ Zamolodchikov, {\em Boundary     
$S$-Matrix and Boundary State in Two-Dimensional Integrable Field     
Theory}, Int.Jour.Mod.Phys.\ {\bf A9} (1994), 3841; {\tt
hep-th/9306002}      
%
\bibitem{corri97} E.\ Corrigan, {\em On duality and reflection factors  
for the sinh-Gordon model}, preprint DTP--97/33, {\tt hep-th/9707235}      
%
\bibitem{ghosh94b} S.\ Ghoshal, {\em Bound State Boundary $S$-Matrix     
of the Sine-Gordon Model}, Int.Jour.Mod.Phys.\ {\bf A9} (1994), 4801;
{\tt hep-th/9310188}     
%
\bibitem{brade90} H.W.\ Braden, E.\ Corrigan, P.E.\ Dorey and R.\ 
Sasaki, {\em Affine Toda Field Theory and Exact $S$-matrices}, 
Nucl.Phys.\ {\bf B338} (1990), 689 
%
\bibitem{gande95} G.M.\ Gandenberger, {\em Exact $S$-matrices for bound    
states of $a_2^{(1)}$ affine Toda solitons}, Nucl.Phys.\ {\bf B449}   
(1995), 375; {\tt hep-th/9501136}    
%
\bibitem{gande95b} G.M.\ Gandenberger and N.J.\ MacKay, {\em Exact
$S$-matrices for $d_{n+1}^{(2)}$ affine Toda solitons and their bound
states}, Nucl.Phys.\ {\bf B457} (1995), 240; {\tt hep-th/9506169}    
%
\bibitem{gande96} G.M.\ Gandenberger, {\em Exact $S$-matrices for
Quantum Affine Toda Solitons and their Bound States}, Ph.D.\ thesis,
University of Cambridge 1996, unpublished;\\  
{\bf Available as postscript file at:  
{\em http://www.damtp.cam.ac.uk/user/hep/publications.html}}  
%
\bibitem{gande98} G.M.\ Gandenberger, {\em Trigonometric $S$-matrices,
Affine Toda Solitons and Supersymmetry}, Int.Jour.Mod.Phys.\ to
appear; {\tt hep-th/9703158}
%
\bibitem{hollo93} T.J.\ Hollowood, {\em Quantizing $Sl(N)$ Solitons and 
the Hecke Algebra}, Int.Jour.Mod.Phys.\ {\bf A8} (1993), 947; {\tt
hep-th/9203076}      
%
\bibitem{deliu98} G.W.\ Delius, {\em Restricting affine Toda field
theory to the half-line}, preprint {\tt hep-th/9807189}
%
\bibitem{arins79} A.E.\ Arinshtein, V.A.\ Fateev and
A.B.\ Zamolodchikov, {\em Quantum $S$-matrix for the
$(1+1)$--dimensional Todd Chain}, Phys.Lett.\ {\bf 87B} (1979), 389
%
\bibitem{sasak93} R.\ Sasaki, {\em Reflection Bootstrap Equations for  
Toda Field Theory}, preprint YITP/U-93-33, {\tt hep-th/9311027}   
%
\bibitem{kim95b} J.D.\ Kim, {\em Boundary Reflection Matrix     
for $ade$ Affine Toda Theory}, Jour.Phys.\ {\bf A29} (1996), 2163;
{\tt hep-th/9506031}       
%
\bibitem{kim96} J.D.\ Kim and Y.\ Yoon, {\em Root Systems and Boundary
Bootstrap}, preprint KAIST/THP-96/701; {\tt hep-th/9603111}      
%
\bibitem{fring94} A.\ Fring and R.\ K{\"o}berle, {\em Factorized
Scattering in the Presence of Reflecting Boundaries}, Nucl.Phys.\ {\bf
B421} (1994), 159; {\tt hep-th/9304141}\\
A.\ Fring and R.\ K{\"o}berle, {\em Affine Toda Field Theory in the
Presence of Reflecting Boundaries}, Nucl.Phys.\ {\bf B419} (1994),
647; {\tt hep-th/9309142}  
%
\bibitem{perki98} M.\ Perkins and P.\ Bowcock, {\em Quantum
corrections to the classical reflection factor in $\ato$ Toda field
theory}, preprint DTP-98/49; {\tt hep-th/9807146} 
%
\bibitem{bowco96} P.\ Bowcock, E.\ Corrigan and R.H.\ Rietdijk, {\em
Background field boundary conditions for affine
Toda field theories}, Nucl.Phys.\ {\bf B465} (1996), 350; {\tt
hep-th/9510071}  
%
\bibitem{fujii95} A.\ Fujii and R.\ Sasaki, {\em Boundary Effects in
Integrable Field Theory on a Half-Line}, Prog.Theor.Phys.\ {\bf 93}
(1995), 1123;  {\tt hep-th/9503083}   
%
\bibitem{deveg93} H.J.\ de Vega and A.\ Gonz\'{a}lez Ruiz, {\em
Boundary $K$-matrices for the six vertex and the $n(2n-1)$ $A_{n-1}$
vertex models}, Jour.Phys.\ {\bf A26} (1993), L519
%
\bibitem{doiko98} A.\ Doikou and R.\ Nepomechie {\em Duality and
quantum--algebra symmetry of the $A_{N-1}^{(1)}$ open spin chain with
diagonal boundary fields}, preprint UMTP--207; {\tt hep-th/9807065}
%
\end{thebibliography}        
}
\end{document}